\documentclass[prd,superscriptaddress,nofootinbib,notitlepage]{revtex4-1}
\usepackage{epsfig,amsmath,amssymb,slashed}
\usepackage{graphicx}
\usepackage{color}
\usepackage{siunitx}
\usepackage{color}
\usepackage{bm}
\usepackage{subfig}

\newcommand{\di}{{\rm d}}
\newcommand{\ii}{i}

\def\wT{{\widehat T}}
\def\wj{{\widehat j}}
\def\wJ{{\widehat J}}

\def\wP{{\widehat P}}

\def\wQ{{\widehat Q}}

\newcommand{\tr}{{\rm tr}}  
\newcommand{\Tr}{{\rm Tr}}  
  
\newcommand{\e}{{\rm e}}

\newcommand{\be}{\begin{equation}}
\newcommand{\ee}{\end{equation}}                                                                               
\def\bea{\begin{eqnarray}}
\def\eea{\end{eqnarray}}      
\begin{document}

\title{Spin polarization of fermions at local equilibrium: Second-order gradient expansion} 

\author{Xin-Li Sheng}
\email{sheng@fi.infn.it}
\affiliation{INFN Sezione di Firenze, Florence, Italy}

\author{Francesco Becattini}
\email{becattini@fi.infn.it}
\affiliation{Universit\`a di Firenze and INFN Sezione di Firenze, Florence, Italy}

\author{Xu-Guang Huang}
\email{huangxuguang@fudan.edu.cn}
\affiliation{Physics Department and Center for Field Theory and Particle Physics, Fudan University, Shanghai 200438, China}
\affiliation{Key Laboratory of Nuclear Physics and Ion-beam Application (MOE), Fudan University, Shanghai 200433, China}
\affiliation{Shanghai Research Center for Theoretical Nuclear Physics, National Natural Science Foundation of China and Fudan University, Shanghai 200438, China}

\author{Zhong-Hua Zhang}
\email{zhzhang22@m.fudan.edu.cn}
\affiliation{Physics Department and Center for Field Theory and Particle Physics, Fudan University, Shanghai 200438, China}

\begin{abstract}
We present a calculation of the spin polarization of spin-1/2 fermions in a relativistic
fluid at local thermodynamic equilibrium at the second order in the gradient expansion,
including second-order derivatives. The second-order derivative terms vanish if the local 
equilibrium hypersurface is the hyperplane $t=const$ in the collision center-of-mass frame. 
However, since the freeze-out hypersurface has a non-trivial space-time structure, these 
terms may result in a non-vanishing contribution to the spin polarization, whose magnitude 
needs to be assessed with numerical computations.   
\end{abstract}

\maketitle

\section{Introduction}

Spin polarization in a relativistic fluid is a subject of major phenomenological relevance in 
relativistic heavy-ion collisions, where the Quark Gluon Plasma (QGP) is formed; for recent 
reviews on spin polarization, see, e.g., \cite{Huang:2020dtn,Gao:2020lxh,Becattini:2024uha}. 

Global and local polarization of $\Lambda$ hyperons have been measured by the STAR collaboration 
at RHIC energies \cite{STAR:2017ckg,STAR:2018gyt,STAR:2019erd}, and by the ALICE collaboration 
at the LHC energy \cite{ALICE:2019onw,ALICE:2021pzu}. The hydrodynamic model with statistical 
hadronic freeze-out proved to be successful in reproducing the data for several observables. 
Global spin polarization in heavy ion collision at high energy was quantitatively and successfully
predicted by this model \cite{Becattini:2015ska,Karpenko:2016jyx,Fu:2020oxj} with the spin-thermal vorticity 
coupling \cite{Becattini:2013fla}, whereas its momentum dependence could not be reproduced \cite{Niida:2018hfw,STAR:2019erd,Becattini:2020ngo}. 
This motivated further theoretical investigations on other possible sources of spin polarization 
in heavy-ion collisions, which led to the finding that also the (thermal) shear tensor and the gradient 
of chemical potentials contribute \cite{Becattini:2021suc,Liu:2021uhn,Liu:2020dxg} at the linear 
order in the gradients of the hydro-thermodynamic fields (see also studies in refs. 
\cite{Fu:2021pok, Yi:2021ryh, Wu:2022mkr, Alzhrani:2022dpi, Weickgenannt:2022zxs}). By 
including the spin polarization induced by the thermal shear tensor, the models can 
reproduce the experimental data \cite{Becattini:2021iol,Palermo:2024tza} with a noteworthy
sensitivity on the initial hydrodynamic conditions as well as on the transport coefficients
 \cite{Yi:2021ryh,Palermo:2024tza}.

In hydrodynamics and non-equilibrium statistical mechanics, any physical observable can
be expressed in terms of the hydro-thermodynamic fields through a so-called gradient expansion
\cite{Kovtun:2012rj,Romatschke:2017ejr}, which provides a good approximation when these fields 
are slowly varying in spacetime. More precisely, the expansion in gradients is appropriate when 
the Knudsen number $K_n$, that is the ratio between the interaction length and the typical scale 
of the inhomogeneity of the fields satisfies $K_n \ll 1$, the so-called hydrodynamic limit 
\footnote{The Knudsen number can be defined as $K_n = \lambda/|\beta/\partial \beta|$ where 
$\lambda$ is the interaction length and $\beta$ the four-temperature field.}. 
However, there are two kinds of terms in the general gradient expansion which are 
conceptually distinct: those stemming from the expansion of the local equilibrium part of the 
density operator, which are non-dissipative, and those stemming from the expansion of the
dissipative part of the density operator, which are associated to the so-called transport 
coefficients \cite{Becattini:2019dxo}, such as shear viscosity. 

It is known that the gradient expansion of the local equilibrium density 
operator gives rise to terms proportional to thermal vorticity tensor $\varpi_{\mu\nu}$, thermal 
shear tensor $\xi_{\mu\nu}$ and gradient of chemical potential for the spin polarization, which 
are first order in the gradient expansion \cite{Becattini:2021suc}. 
In heavy-ion collisions, numerical simulations show that components of these fields are small enough \cite{Becattini:2015ska,Teryaev:2015gxa,Karpenko:2016jyx,Deng:2016gyh,Jiang:2016woz,Li:2017slc,Wei:2018zfb,Deng:2020ygd,Becattini:2021iol} to ensure the validity of the hydrodynamic limit, hence of the gradient expansion, 
up to first order. However, some components, such as the $\varpi_{tz}$ component as shown 
in ref. \cite{Karpenko:2016jyx}, could have significant variations over the freeze-out hypersurface. 
A naive estimation according to ref. \cite{Karpenko:2016jyx} shows that $\partial_x \varpi_{tz}$ 
could be as large as $0.3\,\mathrm{fm}^{-1} \approx 0.05\,\mathrm{GeV}$ at some points on the 
hypersurface. Assuming a typical energy scale $E\approx 1\,\mathrm{GeV}$, we will find that 
$\partial_x \varpi_{tz}/E\approx 0.05$, which is the same order of $\varpi_{tz}$ itself. 
This might be an indication of a sizeable polarization induced by second order terms such as 
$\partial_{\rho} \varpi_{\mu\nu}$ and $\partial_{\rho}\xi_{\mu\nu}$. Nevertheless, we do not 
have any knowledge on the magnitude of second order spin polarizations in the gradient expansion 
due to the lack of theoretical calculations and numerical simulations. In this work, we aim at 
partially filling this gap by analytically calculating, in the linear response approximation,
the spin polarization to the second order gradient of hydrodynamic quantities at local thermodynamic 
equilibrium. On the other hand, we will not delve into the possible dissipative contributions
(i.e. arising from the expansion in the dissipative part of the density operator), which might 
be as important as the non-dissipative contributions but are certainly more difficult to estimate 
numerically because they generally depend on unknown coefficients.

The paper is organized as follows: sec. \ref{sec:Local-equilibrium} features a general discussion 
on the mean values of operators at local equilibrium. In sec. \ref{sec:Wigner-function}, we 
derive the mean value of the Wigner function for spin-1/2 fermions. In sec. \ref{sec:Spin-polarization} 
we calculate the mean spin polarization vector up to second order in gradient expansion. Finally,
section \ref{sec:Summary} presents the summary of this work and the related outlook. 
\subsection*{Notation}

In this paper we use the natural units, with $\hbar=c=K=1$. The Minkowskian metric tensor is 
${\rm diag}(1,-1,-1,-1)$; for the Levi-Civita symbol we use the convention $\epsilon^{0123}=1$. 
We will use the relativistic notation with repeated indices assumed to be saturated. Operators in Hilbert 
space will be denoted by a wide upper hat, e.g. $\widehat H$, except for the Dirac field operator
$\psi (x)$.

\begin{figure}
  \includegraphics[width=0.5\textwidth]{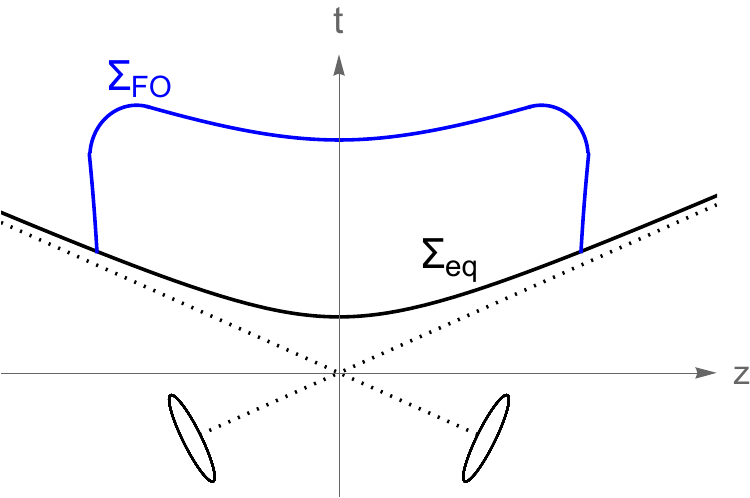}
  \caption{\label{fig:freezeout} 
  Spacetime diagram of a relativistic heavy-ion collision. Here $\Sigma_\text{eq}$ (black line) 
  is the hypersurface where LTE is achieved and $\Sigma_\text{FO}$ (blue line) is the hypersurface where 
  the quark-gluon plasma decouples.}
\end{figure}

\section{Local thermodynamic equilibrium and its expansion} \label{sec:Local-equilibrium}

For a relativistic fluid at Local Thermodynamic Equilibrium (LTE), the density operator is given by \cite{Zubarev:1979afm,VANWEERT1982133,Becattini:2014yxa}
\begin{equation}\label{LTE-rho}
\widehat{\rho}_{\text{LE}}=\frac{1}{Z_\text{LE}}\exp\left[ -\int_\Sigma \di 
\Sigma_\mu \; \left( \widehat{T}^{\mu\nu}(y)\beta_\nu(y)-\frac{1}{2}\widehat{S}^{\mu\nu\lambda}(y)
\Omega_{\nu\lambda}(y)-\widehat{j}^\mu(y)\zeta(y) \right)\right]\,,
\end{equation}
where $Z_{\text{LE}}$ is the normalization factor ensuring $\Tr\,\widehat{\rho}_{\text{LE}}=1$,  
$\beta_\nu=u_\nu/T$  is the four-temperature vector, $\Omega_{\nu\lambda}$ is the spin potential, 
$\zeta=\mu/T$ is the ratio between a chemical potential and temperature, and $y$ is the point 
on the hypersurface $\Sigma$. For heavy ion collisions, we are mostly interested on calculations
at the freeze-out, so the hypersurface is the $\Sigma_\text{FO}$ featured in fig. \ref{fig:freezeout}. 
The formula \eqref{LTE-rho} includes the possibility of a spin potential $\Omega_{\nu\lambda}$ not 
coinciding with thermal vorticity: 
$$
\varpi_{\nu\lambda} = \frac{1}{2} \left(  \partial_\lambda \beta_\nu - \partial_\nu \beta_\lambda \right)\,,
$$
which makes it certainly pseudo-gauge dependent \cite{Becattini:2018duy}. Thus, the operator \eqref{LTE-rho} 
pertains to a particular choice of the pseudo-gauge, for instance the so-called canonical one 
where $\widehat{T}^{\mu\nu}$ and $\widehat{S}^{\mu\nu\lambda}$ are obtained from the Lagrangian 
of the theory by applying the N\"other theorem. 

When calculating the mean value of a local operator $\widehat{O}(x)$ at LTE:
$$
 \langle \widehat{O}(x) \rangle_{\rm LE} = 
 \Tr \left( \widehat{O}(x) \widehat{\rho}_{\text{LE}} \right)
$$
in a relativistic fluid with slowly varying thermodynamic fields $\beta,\zeta$ (small Knudsen
number), it is expected that the above mean value will be mostly determined by the value of those 
fields at the point $x$ while their values far from $x$ will give rise to corrections. One can then 
expand the fields $\beta_\nu(y)$, $\Omega_{\nu\lambda}(y)$, and $\zeta(y)$ in Taylor series around the point 
$x$ where the operator $\widehat{O}$ is to be evaluated, thereby generating a gradient expansion. 
At the leading order, one can approximate:
\begin{equation*}
    \beta_\nu(y)-\beta_\nu(x) \simeq (y-x)^\lambda \partial_\lambda\beta_\nu(x)\,, \qquad
    \qquad \zeta(y)-\zeta(x) \simeq (y-x)^\lambda\partial_\lambda\zeta(x) \,,
\end{equation*}
implying ($\varpi=\Omega$ for the sake of simplicity) \cite{Becattini:2014yxa}:
\be\label{rhoexpand}
\widehat{\rho}_{\text{LE}} \simeq \frac{1}{Z_\text{LE}} \exp\left[ - \beta_\nu(x) \wP^\nu 
+\frac{1}{2} \varpi_{\nu\lambda}(x) \widehat{J}_x^{\nu\lambda} - \frac{1}{2} \xi_{\nu\lambda}(x) 
\widehat Q_x^{\nu\lambda} + \zeta(x) \wQ + \partial_\nu \zeta(x) \widehat D_x^\nu \right]\,,
\ee
where $\xi$ is the thermal shear tensor:
$$
\xi_{\nu\lambda} = \frac{1}{2} \left(  \partial_\lambda \beta_\nu + \partial_\nu \beta_\lambda \right)\,,
$$
$\wP^\mu$ are the four-momentum operators, $\wJ_x^{\lambda\nu}$ are the angular momentum-boost operators with respect to the point $x$, and the operators $\widehat Q_x$ and $\widehat D_x$ read:
\be\label{qandd}
\wQ_x^{\lambda\nu} = \int_\Sigma \di \Sigma_\mu(y) \;\left[ (y-x)^\lambda \wT^{\mu\nu} + 
(y-x)^\nu \wT^{\mu\lambda}\right]\,, 
\qquad\qquad
\widehat D_x^\lambda = \int_\Sigma \di \Sigma_\mu(y) \; (y-x)^\lambda \wj^{\mu}\,.
\ee
The operators $\wQ_x$ and $\widehat D_x$ are responsible for the shear-induced polarization
\cite{Becattini:2021suc,Liu:2021uhn} and spin Hall effect \cite{Liu:2020dxg} respectively
and, unlike $\wP$ and $\wJ$, are dependent on the hypersurface $\Sigma$; this is a crucial
difference because the corrections stemming from $\wQ_x$ and $\widehat D_x$ will, in general,
be dependent on the normal vector field to the hypersurface besides the thermo-hydrodynamic fields.

It is the purpose of this paper to go beyond the leading order Taylor expansion of the thermodynamic
fields and to include the second derivatives of $\beta$ and $\zeta$ in the expansion \eqref{rhoexpand}, 
which will eventually be associated to integral operators of the same sort of $\wQ_x$ and $\widehat D_x$.
To maintain a general viewpoint, it is convenient to recast $\widehat{\rho}_{\text{LE}}$ as:
\begin{equation}\label{LTE-rho-new}
\widehat{\rho}_{\text{LE}}=\frac{1}{Z_\text{LE}}\exp\left[\widehat{A}_x+\widehat{B}_x\right]\,,
\end{equation}
where the operators $\widehat{A}_x$ and $\widehat{B}_x$ are defined as:
\begin{eqnarray} \label{operator-AB}
\widehat{A}_x&\equiv& -\beta_\nu(x)\widehat{P}^\mu+\zeta(x)\widehat{Q}\,, \nonumber\\
\widehat{B}_x&\equiv& -\int_\Sigma \di\Sigma_\mu \; \widehat{T}^{\mu\nu}(y)
\left[\beta_\nu(y)-\beta_\nu(x)\right]+\frac{1}{2}\int_\Sigma \di \Sigma_\mu \; 
\widehat{S}^{\mu\nu\lambda}(y)\Omega_{\nu\lambda}(y) \nonumber\\
&&+\int_\Sigma \di \Sigma_\mu \; \widehat{j}^\mu(y)\left[\zeta(y)-\zeta(x)\right]\,,
\end{eqnarray}
with $\widehat{P}^\nu\equiv\int_\Sigma \di \Sigma_\mu \, \widehat{T}^{\mu\nu}$ being the 
four-momentum operator and $\widehat{Q}\equiv\int_\Sigma \di \Sigma_\mu \widehat{j}^\mu$ 
being the total charge. Due to energy-momentum conservation and current conservation, the 
operator $\widehat{A}_x$ is independent of the hypersurface $\Sigma$ unlike $\widehat{B}_x$.
In eq. \eqref{operator-AB} we have not extracted the spin potential at the point $x$.
Indeed, since the spin potential coincides with the thermal vorticity at global equilibrium, 
we have considered the spin potential $\Omega_{\nu\lambda}(y) \approx \Omega_{\nu\lambda}(x)$ 
as a first order term. In this power counting scheme, the contribution from $\widehat{B}_x$ 
is at least $\mathcal{O}(\partial)$ and therefore is assumed to be much smaller than the contribution 
from $\widehat{A}_x$, allowing us to expand the exponential in (\ref{LTE-rho-new}) up to 
the quadratic term in $\widehat{B}_x$ using the Kubo identity,
\begin{eqnarray*}
\e^{\widehat{A}_x+\widehat{B}_x}&\approx& \e^{\widehat{A}_x}+\int_0^1 \di z \; 
\e^{z\widehat{A}_x}\widehat{B}_x\e^{-z\widehat{A}_x} \nonumber\\
&&+\int_0^1 \di z_1 \int_0^{z_1} \di z_2 \; \e^{z_1\widehat{A}_x}\widehat{B}_x 
 \e^{(z_1-z_2)\widehat{A}_x} \widehat{B}_x \e^{-z_1\widehat{A}_x}+\mathcal{O}(\widehat{B}^3)\,.
\end{eqnarray*}
For the local operator $\widehat{O}(x)$, the mean value at LTE is thus given by:
\begin{eqnarray} \label{Mean_value_O}
\left\langle\widehat{O}(x)\right\rangle_\text{LE}&\approx&\left\langle\widehat{O}(0)\right\rangle_A+
\int_0^1 \di z\;\left\langle\widehat{O}(0),\,\e^{z\widehat{A}_x}\widehat{B}_0\e^{-z\widehat{A}_x}
\right\rangle_{A,C} \nonumber \\
&& + \int_0^1 \di z_1 \int_0^{z_1}\di z_2 \left\langle\widehat{O}(0) ,\, \e^{z_2\widehat{A}_x}
\widehat{B}_0\e^{-z_2\widehat{A}_x},\,\e^{z_1\widehat{A}_x}\widehat{B}_0 \e^{-z_1\widehat{A}_x}
\right\rangle_{A,C}+\mathcal{O}(\widehat{B}^3)\,,
\end{eqnarray}
where $\left\langle \;\; \right\rangle_A$ denotes the mean value at the homogeneous global 
thermodynamic equilibrium described by $\widehat{A}$:
\begin{equation*}
\left\langle\widehat{O}\right\rangle_A\equiv\frac{\text{Tr}\left[\widehat{O}\,\e^{\widehat{A}_x}\right]}
{\text{Tr}\left[\e^{\widehat{A}_x}\right]}\,.
\end{equation*}
In eq. \eqref{Mean_value_O} we have shifted the operator $\widehat{O}(x)$ by means of 
the identity $\widehat{O}(0) \equiv \e^{-\ii x \cdot \widehat{P}}\widehat{O}(x) \e^{\ii x\cdot\widehat{P}}$ 
and $\widehat{B}_x$ likewise. The connected mean values are defined as:
\begin{eqnarray}\label{connected-mean}
\left\langle\widehat{O}_1,\,\widehat{O}_2\right\rangle_{A,C}&\equiv& 
\left\langle\widehat{O}_1\widehat{O}_2\right\rangle_A-\left\langle\widehat{O}_1\right\rangle_A
\left\langle\widehat{O}_2\right\rangle_A\,,\nonumber\\
\left\langle\widehat{O}_1,\,\widehat{O}_2,\,\widehat{O}_3\right\rangle_{A,C}&\equiv& \left\langle\widehat{O}_1\widehat{O}_2\widehat{O}_3\right\rangle_A -\
\left\langle\widehat{O}_1\widehat{O}_2\right\rangle_A \left\langle\widehat{O}_3\right\rangle_A -\left\langle\widehat{O}_2\widehat{O}_3\right\rangle_A \left\langle\widehat{O}_1\right\rangle_A\nonumber\\
&& -\left\langle\widehat{O}_1\widehat{O}_3\right\rangle_A \left\langle\widehat{O}_2\right\rangle_A+
2\left\langle\widehat{O}_1\right\rangle_A\left\langle\widehat{O}_2\right\rangle_A 
\left\langle\widehat{O}_3\right\rangle_A\,,
\end{eqnarray}
for arbitrary operators $\widehat{O}_{1,2,3}$.

We now focus on free spin-1/2 fermions, for which the canonical energy-stress tensor, canonical spin 
tensor, and the current are given by \cite{Itzykson:1980rh},
\begin{eqnarray}\label{canonical-T-S}
\widehat{T}^{\mu\nu}(x)&=&\frac{i}{2}\overline{\psi}(x)\gamma^\mu(\overrightarrow{\partial}^\nu
-\overleftarrow{\partial}^\nu)\psi(x)\,,\nonumber\\
\widehat{S}^{\mu\nu\lambda}(x)&=&-\frac{1}{2}\epsilon^{\mu\nu\lambda\rho}\overline{\psi}(x)
\gamma_\rho\gamma^5\psi(x)\,, \nonumber\\
\widehat{j}^\mu(x)&=&\overline{\psi}(x)\gamma^\mu\psi(x)\,,
\end{eqnarray}
where $\psi(x)$ is the Dirac field.
By substituting (\ref{canonical-T-S}) into eq. (\ref{operator-AB}) and taking the 4D Fourier transforms, 
it is possible to express $\widehat{B}_0$ as follows:
\begin{equation}\label{B-projected}
  \widehat{B}_0=\e^{-\ii x\cdot\widehat{P}}\widehat{B}_x \e^{\ii x\cdot\widehat{P}}
  =\frac{1}{(2\pi)^5} \int \di^4 k_1 \di^4 k_2 \; \overline{\psi}(k_2) \mathcal{B}(k_2,k_1)
  \psi(k_1)\,,
\end{equation}
where $\psi(k)$ is the Fourier component of the Dirac field,
\begin{equation} \label{Fourier_Dirac}
 \psi(k)\equiv \int \di^4 x \; \psi(x) \e^{\ii k\cdot x},\qquad 
 \overline{\psi}(k)\equiv \psi^\dagger(k)\gamma^0 = \int \di^4x \; \overline{\psi}(x)\e^{-\ii k\cdot x}\,.
\end{equation}
Since $\widehat{B}_0$ is Hermitian, $\mathcal{B}(k_2,k_1)$ is a $4\times 4$ Dirac matrix 
fulfilling $\gamma^0 \mathcal{B}^\dagger(k_2,k_1)\gamma^0=\mathcal{B}(k_1,k_2)$, whose
explicit form reads:
\begin{equation}\label{function_B_Sigma}
\mathcal{B}(k_2,k_1)=-\frac{1}{(2\pi)^3}\int_\Sigma \di \Sigma_\mu \; \e^{-\ii q\cdot(y-x)}
\bigg\{\gamma^\mu k^\nu\left[\beta_\nu(y)-\beta_\nu(x)\right]
+\frac{1}{4}\epsilon^{\mu\nu\lambda\sigma}\gamma_\sigma\gamma^5 
\Omega_{\nu\lambda}(y) -\gamma^\mu \left[\zeta(y)-\zeta(x)\right]\bigg\}\,,
\end{equation}
where $k\equiv(k_1+k_2)/2$ is the average momentum and $q\equiv k_1-k_2$ is the relative 
momentum, respectively. The fields $\beta_\nu(y)$, $\Omega_{\nu\lambda}(y)$, and $\zeta(y)$ 
can be expressed as Taylor series with respect to $y-x$. For instance:
\begin{equation*}
\beta_\nu(y)=\beta_\nu(x)+\sum_{n=1}^\infty \frac{1}{n!}\left[\prod_{j=1}^n\left(y^{\lambda_j}-x^{\lambda_j}\right)\partial_{\lambda_j}\right]\beta_\nu \Big|_{y=x},
\end{equation*}
where $\partial_{\lambda_j}$ acts on $\beta_\nu$. Then:
\begin{align*}
& \e^{-\ii q\cdot(y-x)}\left[ \beta_\nu(y) -\beta_\nu(x)\right] = \e^{-\ii q\cdot(y-x)}
\sum_{n=1}^\infty \frac{1}{n!}\left[\prod_{j=1}^n\left(y^{\lambda_j}-x^{\lambda_j}\right)
\partial_{\lambda_j}\right]\beta_\nu \Big|_{y=x} \\
& = \sum_{n=1}^\infty \frac{1}{n!}\left[\prod_{j=1}^n \partial_{\lambda_j}\right]\beta_\nu \Big|_{y=x}
\left[\prod_{j=1}^n \ii \frac{\partial}{\partial_{q_j}} \right] \e^{-\ii q\cdot(y-x)}
\equiv [\beta_\nu (x + \ii \partial_q) - \beta_\nu(x)]\left( \e^{-\ii q\cdot(y-x)} \right)\,.
\end{align*}
The other terms in \eqref{function_B_Sigma} involving $\Omega$ and $\zeta$ can be 
expressed likewise. Finally, all the functions $\beta(x + \ii \partial_q)$ etc. can
be moved out of the integral, so that eq. \eqref{function_B_Sigma} turns into:
\begin{equation}\label{function_B_Sigma2}
\mathcal{B}(k_2,k_1)= -\frac{1}{(2\pi)^3} \bigg\{\gamma^\mu k^\nu
\left[\beta_\nu(x+\ii \partial_q)-\beta_\nu(x)\right]
+\frac{1}{4}\epsilon^{\mu\nu\lambda\sigma}\gamma_\sigma\gamma^5 
\Omega_{\nu\lambda}(x+\ii \partial_q) -\gamma^\mu 
\left[\zeta(x + \ii \partial_q)-\zeta(x)\right]\bigg\}
\int_\Sigma \di \Sigma_\mu \; \e^{-\ii q\cdot(y-x)}\,,
\end{equation}
where it is understood that the term within the curl brackets acts on the integral 
over $\Sigma$. Now, with the help of the Gauss's theorem, that integral can be converted 
to an integral over a portion of a hyperplanes $\Sigma_B$ with constant normal vector 
$\hat{t}^\mu=(1,0,0,0)$ and an integral over a 4D region $\Omega_B$  enclosed by 
two space-like hyperplanes and a part of the hypersurface $\Sigma$ (see fig. \ref{fig:freezeout2} in
Appendix \ref{app:hyp}). Assuming that the space extent of $\Sigma_B$  and $\Omega_B$ 
are sufficiently large, we obtain (see Appendix \ref{app:hyp} for the derivation): 
\begin{eqnarray}\label{eq:hypersurface-int}
\int_\Sigma \di \Sigma_\mu \;  \e^{-\ii q\cdot(y-x)}&\simeq& (2\pi)^3\left[\hat{t}_\mu 
\e^{-\ii q^0(y_B^0-x^0)}
-\ii q_\mu \int_{y_B^0}^{y_\Sigma^0(q)}\di y^0 \; \e^{-\ii q^0(y^0-x^0)}\right] \delta^{3}
({\bf q})\nonumber\\
&=&(2\pi)^3 \e^{-\ii q^0(y_\Sigma^0(q)-x^0)}\delta^{3}({\bf q}) \, \hat{t}_\mu\,,
\end{eqnarray}
where $y_B^0$ denotes the constant time for $\Sigma_B$ and $y_\Sigma^0(q)$ denotes a
costant time, dependent on $q$, which can be considered as an average time for the space-like
part of the hypersurface $\Sigma$ (see again Appendix \ref{app:hyp} for the precise 
definition).
By plugging eq. (\ref{eq:hypersurface-int}) into eq. (\ref{function_B_Sigma2}), 
we can finally express $\mathcal{B}(k_2,k_1)$ as follows:
\begin{eqnarray} \label{function_B_k2k1}
\mathcal{B}(k_2,k_1)& \simeq&-\hat{t}_\mu\bigg\{\gamma^\mu k^\nu\left[\beta_\nu(x+i\partial_q)-
\beta_\nu(x)\right]+\frac{1}{4}\epsilon^{\mu\nu\lambda\rho}\gamma_\rho\gamma^5
\Omega_{\nu\lambda}(x+i\partial_q)\bigg. \nonumber \\
&& \bigg.-\gamma^\mu\left[\zeta(x+i\partial_q)-\zeta(x)\right]\bigg\}
\left[\e^{-\ii q^0(y_\Sigma^0(q)-x^0)}\delta^{3}({\bf q})\right]\,,
\end{eqnarray}
with the same proviso as for eq. \eqref{function_B_Sigma2}.

\section{The Wigner function at local equilibrium} \label{sec:Wigner-function}

The general formula to calculate the mean spin vector of spin 1/2 fermions reads \cite{Becattini:2020sww}:
\begin{equation}\label{spindef}
S^\mu(p)=\frac{1}{2}\frac{\int_\Sigma \di \Sigma\cdot p_+\,\tr[\gamma^\mu \gamma^5 W(x,p)]}
{\int_\Sigma \di \Sigma\cdot p_+\,\tr[W(x,p)]}\,,
\end{equation}
where $\tr$ stands for the trace over the spinorial indices and $p_+^\mu\equiv p^\mu\theta(p^0)$ 
selects the positive-energy part of the Wigner function, removing antiparticle contribution. 
The covariant Wigner function is defined as \cite{Heinz:1983nx,Vasak:1987um}:
\begin{equation}\label{Wigner-definition}
W_{ab}(x,p)\equiv\frac{1}{(2\pi)^4}\int \di^4 y\; \e^{-\ii p\cdot y}\left\langle\overline\psi_b
\left(x+\frac{y}{2}\right)\psi_a \left(x-\frac{y}{2}\right)\right\rangle\,,
\end{equation}
where $a,b$ are spinorial Dirac indices and $\left\langle(\cdots)\right\rangle\equiv
\text{Tr}[\widehat{\rho}(\cdots)]$ denotes the mean value with a specific density matrix. 
It should be stressed that, even though the argument $p$ of the Wigner function is not 
necessarily on-shell, in the formula \eqref{spindef}, for a free field, $p$ becomes 
an on-shell four-vector.
The reason is that, for the Dirac free field, the divergence of the integrand 
in equation \eqref{spindef} vanishes, that is the Wigner function, either restricted to 
particles or not, fulfills the equation $p^\mu \partial_\mu W(x,p)=0$. This condition 
makes it possible to choose an arbitrary hypersurface to perform integration e.g. in the equation 
\eqref{spindef} or in any other similar expression and to show, by means of an explicit calculation 
for a free field, that only on-shell four-vectors $p$ contribute to the integral \cite{Becattini:2020sww}.

The goal of this section is the calculation of the Wigner function at LTE by using the 
expansion in eq. (\ref{Mean_value_O}). For this purpose, we first make use of the 
Fourier decomposition in eq. (\ref{Fourier_Dirac}) to convert the definition 
(\ref{Wigner-definition}) to the momentum space:
\begin{equation} \label{eq:Wigner-momentum}
W_{ab}(x,p)=\frac{1}{(2\pi)^8}\int \di^4k_1 \di^4k_2 \,\delta^{4}\left(p-\frac{k_1+k_2}{2}\right)
\e^{-i(k_1-k_2)\cdot x}\left\langle\overline{\psi}_b(k_2)\psi_a(k_1)\right\rangle\,,
\end{equation}
then we apply eq. (\ref{Mean_value_O}) by setting $\widehat{O}=\overline{\psi}_b(k_2)\psi_a(k_1)$. 
The result can be in general written as a series in powers of  $\widehat{B}$:
\begin{equation}\label{B-expansion}
W_{\text{LE}}(x,p)=W_0 (x,p)+W_{\text{lin}}(x,p)+W_{\text{quad}}(x,p)+\mathcal{O}(\widehat{B}^3)\,.
\end{equation}
We are now going to determine the form of the Wigner function up to the quadratic term.

\subsection{Main term}

The $\widehat{B}$-independent part of the Wigner function is given by:
\begin{equation} \label{W-indep}
W_{0,ab}(x,p)= \frac{1}{(2\pi)^8}\int \di^4 q \; \e^{-iq\cdot x} 
 \left\langle\overline\psi_b(p_2)\psi_a(p_1)\right\rangle_{A}\,,
\end{equation}
where 
\be\label{pdefs}
 p\equiv(p_1+p_2)/2\,, \qquad q\equiv p_1-p_2 \implies p_{1,2}\equiv p\pm q/2 
\ee
and $\widehat{A}_x$ is given by eq. \eqref{operator-AB}. Noting that the field operators 
in coordinate space obey the equal-time anticommutating relation $\left\{\psi_a(t,{\bf x}),
\psi_b^\dagger(t,{\bf x}^\prime)\right\}= \ii \delta^{3}({\bf x}-{\bf x}^\prime)\delta_{ab}$,
one can readily derive the commutators $\left[\psi_a(p),\widehat{P}^\mu\right]=p^\mu\psi_a(p)$ 
and $\left[\psi_a(p),\widehat{Q}\right]=\psi_a(p)$. It then follows that the expectation value
in eq. (\ref{W-indep}) is given by 
\begin{equation} \label{two-field-corr}
\left\langle\overline\psi_b(p_2)\psi_a(p_1)\right\rangle_{A}=n_F(x,p_1)
\left\{\overline\psi_b(p_2),\psi_a(p_1)\right\}\,,
\end{equation}
where $n_F$  is the Fermi-Dirac distribution parameterized by local variables 
$\beta^\mu(x)$ and $\zeta(x)$,
\begin{equation} \label{eq:Fermi-Dirac-dis}
n_F(x,p)\equiv\frac{1}{1+\exp\left[\beta(x)\cdot p-\zeta(x)\right]}\,.
\end{equation}
By quantizing the Dirac field in a standard way, one can show that the anticommutator in 
eq. \eqref{two-field-corr} reads:
\begin{equation}\label{eq:field-anticomm}
\left\{\overline\psi_b(p_2),\psi_a(p_1)\right\}=(2\pi)^5\delta(p^2-m^2)\text{sgn}(p^0) 
(\slashed{p}+m)_{ab}\delta^{4}(q)\,,
\end{equation}
where $\text{sgn}$ is the sign function. Substituting eqs. (\ref{two-field-corr}) and 
(\ref{eq:field-anticomm}) into eq. (\ref{W-indep}), we obtain
\begin{equation} \label{W-indep-final}
W_\text{0}(x,p)=\frac{\delta(p^2-m^2)\text{sgn}(p^0)}{(2\pi)^3}(\slashed{p}+m)n_F(x,p)\,,
\end{equation}
which is exactly the known Wigner function for free fermions 
\cite{Vasak:1987um,Fang:2016vpj,Weickgenannt:2019dks} at global thermodynamic equilibrium. 
We note that $p^0$ in eq. (\ref{W-indep-final}) could be either positive or negative, 
indicating that $W_\text{0}$ contains contributions from both particles and antiparticles. 
We also find that $W_\text{0}$ does not have axial-vector components, indicating 
that particles are not polarized at this order. 

\subsection{Linear term}

The linear term in the expansion of the LTE density operator reads:
\begin{eqnarray}\label{Wlin}
W_{\text{lin},ab}(x,p)&=&\frac{1}{(2\pi)^{13}}\int \di^4 k_1\di^4k_2\di^4k_3\di^4k_4 \; 
\delta^{4}\left(p-\frac{k_1+k_2}{2}\right)\mathcal{B}_{cd}(k_4,k_3) \nonumber\\
&&\times \int_0^1 \di z\; \e^{z\beta\cdot(k_3-k_4)}
\left\langle\overline{\psi}_b(k_2)\psi_a(k_1),\,\overline{\psi}_d(k_4)\psi_c(k_3)\right\rangle_{A,C}\,,
\end{eqnarray}
where $a,b,c,d$ are spinorial Dirac indices and the connected mean value is defined 
in eq. (\ref{connected-mean}). One can check that:
\begin{equation}\label{four-field-corr}
\left\langle\overline{\psi}_b(k_2)\psi_a(k_1),\,\overline{\psi}_d(k_4)\psi_c(k_3)\right\rangle_{A,C}
=n_F(x,k_2)\left[1-n_F(x,k_1)\right]\left\{\overline{\psi}_b(k_2),\psi_c(k_3)\right\}
\left\{\overline{\psi}_d(k_4),\psi_a(k_1)\right\}\,,
\end{equation}
where the anticommutators are given by eq. (\ref{eq:field-anticomm}). Substituting 
eqs. (\ref{eq:field-anticomm}) and (\ref{four-field-corr}) into \eqref{Wlin}, we obtain:
\begin{eqnarray} \label{W-linear-final}
W_\text{lin}(x,p)&=&\frac{1}{(2\pi)^3}\,\int \di^4q \,\delta(p_1^2-m^2)\delta(p_2^2-m^2)
\text{sgn}(p_1^0)\text{sgn}(p_2^0)\nonumber\\
&&\times n_F(x,p_1)\left[1-n_F(x,p_2)\right]\int_0^1 \di z\; \e^{z\beta\cdot q}
(\slashed{p}_2+m)\mathcal{B}(p_2,p_1)(\slashed{p}_1+m)\,,
\end{eqnarray}
where $p_1,p_2,p,q$ are like in eq. \eqref{pdefs}.

\subsection{Quadratic term}

By applying eq. (\ref{Mean_value_O}) to the Wigner function in eq. (\ref{eq:Wigner-momentum}) 
and expressing $\widehat{B}$ in terms of the Dirac field operators as shown in the
eq. (\ref{B-projected}), one can derive the quadratic term in eq. \eqref{B-expansion}:
\begin{eqnarray}\label{eq:W-quad}
W_{\text{quad},ab}(x,p)&=&\frac{1}{(2\pi)^{18}}\int \di^4p_1\di^4p_2\di^4k_1\di^4k_2\di^4k_3\di^4k_4 
\; \delta^{4}\left(p-\frac{p_1+p_2}{2}\right) \nonumber\\
&&\times\int_0^1\di z_1 \int_0^{z_1} \di z_2\; \e^{z_2\beta\cdot(k_1-k_2)+z_1\beta\cdot(k_3-k_4)}
\mathcal{B}_{cd}(k_2,k_1)\mathcal{B}_{ef}(k_4,k_3)\nonumber\\
&&\times \left\langle\overline{\psi}_b(p_2)\psi_a(p_1),\,\overline{\psi}_c(k_2)\psi_d(k_1),\,
\overline{\psi}_e(k_4)\psi_f(k_3)\right\rangle_{A,C}\,,
\end{eqnarray}
where the subscripts $a,b,c,d,e,f$ are spinorial indices. The connected mean value turns out
to be:
\begin{eqnarray*}
&&\left\langle\overline{\psi}_b(p_2)\psi_a(p_1),\,\overline{\psi}_c(k_2)\psi_d(k_1),\,
\overline{\psi}_e(k_4)\psi_f(k_3)\right\rangle_{A,C} \nonumber \\
&=& (1-n_F(x,p_1)) n_F(x,p_2) \left[ (1-n_F(x,k_1))\left\{\overline{\psi}_b(p_2),\psi_f(k_3)\right\}
\left\{\overline{\psi}_c(k_2),\psi_a(p_1)\right\}\left\{\overline{\psi}_e(k_4),\psi_d(k_1)\right\} 
\right.\nonumber\\
&&-\left.n_F(x,k_2)\left\{\overline{\psi}_b(p_2),\psi_d(k_1)\right\}
\left\{\overline{\psi}_c(k_2),\psi_f(k_3)\right\}\left\{\overline{\psi}_e(k_4),\psi_a(p_1)\right\}\right]\,,
\end{eqnarray*}
where the Fermi-Dirac distributions are defined in (\ref{eq:Fermi-Dirac-dis}) and the 
anticommutators are given in eq. (\ref{eq:field-anticomm}). After completing some momentum 
integrals, the equation (\ref{eq:W-quad}) simplifies to the following form:
\begin{eqnarray}\label{eq:W-quad-final}
W_\text{quad}(x,p)&=&\frac{1}{(2\pi)^3}\int \di^4q \, \di^4q^\prime\; \delta(p_1^2-m^2)
\delta(p_2^2-m^2)\delta(p_3^2-m^2) \text{sgn}(p_1^0)\text{sgn}(p_2^0)\text{sgn}(p_3^0)n_F(p_1)[1-n_F(p_3)]
\nonumber\\
&&\times\int_0^1\di z_1\int_0^{z_1}\di z_2 \left\{-\e^{z_2\beta\cdot q+z_1\beta\cdot q^\prime}n_F(p_2) + 
\e^{z_2\beta\cdot q^\prime+z_1\beta\cdot q} [1-n_F(p_2)]\right\}  \nonumber \\
&& \times (\slashed{p}_3+m)\mathcal{B}(p_3,p_2)(\slashed{p}_2+m)\mathcal{B}(p_2,p_1)(\slashed{p}_1+m)\,,
\end{eqnarray}
where we have defined $p_1\equiv p+(q+q^\prime)/2$, $p_2\equiv p+(-q+q^\prime)/2$, 
and $p_3\equiv p-(q+q^\prime)/2$.

\section{Spin polarization at local equilibrium}\label{sec:Spin-polarization}

We can now use the expansion \eqref{B-expansion} in eq. \eqref{spindef} to
determine the spin polarization vector. First, though, we will rearrange the terms of the 
expansion \eqref{B-expansion} in increasing order of gradients:
\begin{equation} \label{W-LE}
W_\text{LE}(x,p)=W^{(0)}(x,p)+W^{(1)}(x,p)+W^{(2)}(x,p)+\mathcal{O}(\partial^3)\,.
\end{equation}
where
\begin{equation}\label{wdecomp}
W^{(0)}=W_0\,,\ \ \ \ W^{(1)}=W_\text{lin}^{(1)}\,,\ \ \ \ W^{(2)}=W_\text{lin}^{(2)}+W_\text{quad}\,,
\end{equation}
and the superscripts denote their orders in the gradient expansion. By first order in the gradient
expansion we mean terms including $\partial_\mu\beta_\nu(x)$, $\partial_\mu \zeta(x)$, and 
$\Omega_{\mu\nu}(x)$, while for second order we mean second order derivatives such as 
$\partial_\mu\partial_\lambda\beta_\nu(x)$, $\partial_\mu\partial_\lambda\zeta(x)$, 
$\partial_\lambda\Omega_{\mu\nu}(x)$ as well as quadratic terms in the first order derivatives 
such as $[\partial_\mu\beta_\nu(x)][\partial_\lambda\beta_\sigma(x)]$, $[\partial_\mu\zeta(x)]
[\partial_\nu\zeta(x)]$. Hence, up to second order in gradient, the spin polarization is given by:
\begin{eqnarray}\label{Spin-polarization}
S^{(0)\mu}(p)&=&0\,, \nonumber\\
S^{(1)\mu}(p)&=&\frac{\int \di\Sigma\cdot p_+\,\tr\left[\gamma^\mu \gamma^5 W^{(1)}(x,p)\right]}
{2\int \di\Sigma\cdot p_+\,\tr[W^{(0)}(x,p)]}\,,\nonumber\\
S^{(2)\mu}(p)&=&\frac{\int \di\Sigma\cdot p_+\,\tr\left[\gamma^\mu \gamma^5 W^{(2)}(x,p)\right]}
{2\int \di\Sigma\cdot p_+\,\tr[W^{(0)}(x,p)]}-\frac{\left\{\int \di\Sigma\cdot p_+\,
\tr\left[\gamma^\mu \gamma^5 W^{(1)}(x,p)\right]\right\}\left\{\int \di\Sigma\cdot p_+\,
\tr[W^{(1)}(x,p)]\right\}}{2\left\{\int \di\Sigma\cdot p_+\,\tr[W^{(0)}(x,p)]\right\}^2}\,, 
\end{eqnarray}
where the zeroth order part $S^{(0)\mu}(p)$ vanishes because $\tr[\gamma^\mu\gamma^5W^{(0)}]=0$.
Also, the denominator in eq. \eqref{Spin-polarization} can be straightforwardly 
obtained from eq. \eqref{W-indep-final}:
\begin{equation} \label{traceW0}
\tr \left[ W_\text{0}(x,p) \right] =4m \frac{\delta(p^2-m^2)\text{sgn}(p^0)}{(2\pi)^3}  n_F(x,p) \; .
\end{equation}
%
\subsection{First order in the gradient expansion}

At the first order in the gradient expansion, the Wigner function is obtained by taking
the first-order term in the expansion of $\mathcal{B}$ in eq. \eqref{function_B_k2k1}:
\be\label{eq:B-first-order}
\mathcal{B}^{(1)}(p_2,p_1)=-\hat{t}_\mu\bigg\{\gamma^\mu 
\left[p^\nu\partial_\lambda\beta_\nu(x)-\partial_\lambda \zeta(x)\right]i\partial_q^\lambda+\frac{1}{4}\epsilon^{\mu\nu\lambda\rho}\gamma_\rho\gamma^5\Omega_{\nu\lambda}(x)\bigg\} 
\left[\e^{-\ii q^0(y_\Sigma^0-x^0)}\delta^{3}({\bf q})\right]\,
\ee
and plugging it into eq. (\ref{W-linear-final}). The result reads:
\be\label{W-first-order}
W^{(1)}(x,p)=\frac{1}{(2\pi)^3}\int \di^4q\,\delta(p_1^2-m^2)\delta(p_2^2-m^2)
\text{sgn}(p_1^0)\text{sgn}(p_2^0) \left[f_A^{\lambda}(p_2,p_1)i\partial^q_\lambda+f_B(p_2,p_1)\right]
\left[\e^{-\ii q^0(y_\Sigma^0-x^0)}\delta^{3}({\bf q})\right]\,.
\ee
where the auxiliary functions $f_{A,B}$ are defined as
\begin{eqnarray*}
f_A^{\lambda}(p_2,p_1)&\equiv&-\hat{t}_\mu\left\{\partial^\lambda
\left[p_\nu\beta^\nu(x)-\zeta(x)\right]\right\} (\slashed{p}_2+m)
\gamma^\mu(\slashed{p}_1+m) n_F(x,p_1)\left[1-n_F(x,p_2)\right]
 \int_0^1 \di z\; \e^{z\beta\cdot q}\,,\nonumber\\
f_B(p_2,p_1)&\equiv&-\frac{1}{4}\hat{t}_\mu\epsilon^{\mu\nu\lambda\sigma}\Omega_{\nu\lambda} 
(\slashed{p}_2+m)\gamma_\sigma\gamma^5 (\slashed{p}_1+m) n_F(x,p_1)\left[1-n_F(x,p_2)\right]
\int_0^1 \di z\, \e^{z\beta\cdot q}\,.
\end{eqnarray*}
and $p,q,p_1,p_2$ are defined like in eq. \eqref{pdefs}.
Noting that the product of the two delta functions can be rewritten as:
\begin{equation}\label{deltaprod}
\delta(p_1^2-m^2)\delta(p_2^2-m^2)=
\frac{1}{2|p^0|}\delta\left(p^2+\frac{q^2}{4}-m^2\right)
\delta\left(q^0-\frac{{\bf p}\cdot{\bf q}}{p^0}\right)\,,
\end{equation}
the integral in $\di^4 q$ in eq. (\ref{W-first-order}) can be carried out by parts, so
to obtain:
\begin{equation}\label{eq:W-first-order}
W^{(1)}(x,p)=\frac{\delta(p^2-m^2)}{(2\pi)^3 2|p^0|}\lim_{q^\mu\rightarrow0}
\left[D_\lambda^q f_A^{\lambda}(p_2,p_1)+f_B(p_2,p_1)\right]\,,
\end{equation}
where the operator $D_\lambda^q$ is defined as:
\begin{equation} \label{operator-D}
D_\lambda^q\equiv -i\partial_\lambda^q +\frac{p_\lambda}{p^0}
\left(i\partial_0^q+y^0_\Sigma(q)-x^0\right)\,.
\end{equation}
Somewhat lengthy but simple calculations involving the trace of multiple $\gamma$ matrices 
give the scalar and axial-vector components of the Wigner function at this order:
\begin{eqnarray}\label{Wigner-components-1}
\tr\left[W^{(1)}(x,p)\right]&=&-\frac{\delta(p^2-m^2)}
{(2\pi)^3|p^0|}n_F(x,p)\left[1-n_F(x,p)\right](y_\Sigma^0-x^0) 
4m p^\lambda\partial_\lambda[p^\sigma\beta_\sigma(x)-\zeta(x)]\,,\nonumber\\
\tr\left[\gamma^\mu \gamma^5 W^{(1)}(x,p)\right]&=&-\frac{\delta(p^2-m^2)}{(2\pi)^3|p^0|}
n_F(x,p)\left[1-n_F(x,p)\right]\nonumber\\
&&\times \left\{2\epsilon^{\mu\nu\rho\lambda}p_\nu\hat{t}_\rho\partial_\lambda
\left[p^\tau\beta_\tau(x)-\zeta(x)\right]+(p^\mu p_\tau-g^\mu_\tau m^2)
\hat{t}_\rho\epsilon^{\rho\nu\lambda\tau}\Omega_{\nu\lambda}(x)\right\}\,.
\end{eqnarray}
Substituting the second equation of \eqref{Wigner-components-1} and eq. \eqref{traceW0} 
in eq. (\ref{Spin-polarization}), the polarization at first order in gradient
expansion can be obtained:
\begin{eqnarray} \label{eq:first-order-polarization}
S^{(1)\mu}(p)&=&-\frac{1}{8mN}\int \di\Sigma\cdot p_+\,n_F(x,p)\left[1-n_F(x,p)\right] \nonumber\\
&&\times\left\{\epsilon^{\mu\nu\lambda\sigma}\Omega_{\nu\lambda}p_\sigma-
\frac{2}{p \cdot \hat t}\hat{t}_\nu\epsilon^{\mu\nu\lambda\sigma}p_\lambda\left[\left(\xi_{\sigma\rho}
+\Omega_{\sigma\rho}-\varpi_{\sigma\rho}\right)p^\rho-\partial_\sigma\zeta\right]\right\}\,,
\end{eqnarray}
where 
\be\label{norm}
 N\equiv \int \di\Sigma\cdot p_+\,n_F(x,p)
\ee
and $\xi_{\sigma\rho}\equiv(\partial_\sigma\beta_\rho+\partial_\rho\beta_\sigma)/2$ is the thermal 
shear tensor. The first term reduces to the polarization induced by thermal vorticity when 
$\Omega_{\nu\lambda}=\varpi_{\nu\lambda}$, which is the major contribution to the $\Lambda$ 
hyperons' global polarization. The contribution proportional to $\xi_{\sigma\rho}$ is the 
shear-induced polarization that coincides with the result of ref. \cite{Becattini:2021suc} 
and, essentially, with that of ref. \cite{Liu:2021nyg}\footnote{In ref. \cite{Liu:2021nyg} 
the normalization of the mean spin polarization vector slightly differs from the most commonly 
used.}. The contribution from $\partial_\sigma\zeta$ is the so-called spin Hall effect 
\cite{Liu:2020dxg} and again, our result essentially coincides with the result reported 
in \cite{Liu:2021nyg}. Finally, the contribution from the difference between spin potential 
and thermal vorticity coincides with the calculation in ref. \cite{Buzzegoli:2021wlg} and,
essentially, with that in ref. \cite{Liu:2021nyg}.

A key feature of eq. \eqref{eq:first-order-polarization} is its dependence
on the vector $\hat t$ which is the direction of time in the centre-of-mass frame of
the colliding nuclei (see fig. \ref{fig:freezeout}). As it was pointed out in ref. 
\cite{Becattini:2021suc}, the appearance of this vector stems from the essential 
dependence of the operator $\wQ^{\lambda\nu}_x$ in eq. \eqref{qandd} on the hypersurface 
$\Sigma$ (unlike $\wJ^{\lambda\nu}_x$); in this respect, $\hat t$ can be considered as
the average normal vector to the hypersurface $\Sigma$.

\subsection{Second order in the gradient expansion}

At the second order in the gradient expansion, the Wigner function contains both linear 
and quadratic contributions, as expressed by eqs.  
\eqref{wdecomp} and \eqref{Spin-polarization}. We first focus on $W_\text{lin}^{(2)}$, which is obtained by plugging the 
second order term of the expansion of eq. \eqref{function_B_k2k1}:
$$
\mathcal{B}^{(2)}(p_2,p_1)=
\hat{t}_\mu\bigg\{\frac12\gamma^\mu \left[p^\nu\partial_\lambda\partial_\sigma\beta_\nu(x)
-\partial_\lambda \partial_\sigma\zeta(x)\right]\partial_q^\lambda\partial_q^\sigma+\frac{i}{4}\epsilon^{\mu\nu\lambda\rho}\gamma_\rho\gamma^5\left[\partial_\sigma
\Omega_{\nu\lambda}(x)\right]\partial_q^\sigma\bigg\}\left[\e^{-\ii q^0(y_\Sigma^0-x^0)}
\delta^{3}({\bf q})\right]
$$
into $W_\text{lin}$ in eq. (\ref{W-linear-final}). The result reads (see Appendix \ref{app:w2lin}):
\begin{eqnarray} \label{W-second-order}
W_\text{lin}^{(2)}(x,p)&=&\frac{1}{(2\pi)^3}\int \di^4q\,\delta(p_1^2-m^2)
\delta(p_2^2-m^2)\text{sgn}(p_1^0)\text{sgn}(p_2^0) \nonumber\\
&&\times\left[f_C^{\lambda\sigma}(p_2,p_1)\partial^q_\lambda\partial^q_\sigma+
f_D^{\sigma}(p_2,p_1)i\partial_\sigma^q\right]\left[\e^{-\ii q^0(y_\Sigma^0-x^0)}
\delta^{3}({\bf q})\right]\,,
\end{eqnarray}
where the auxiliary functions $f_{C,D}$ are defined as:
\begin{eqnarray} \label{functions_f3_f4}
f_C^{\lambda\sigma}(p_2,p_1)&\equiv&\frac{1}{2}\hat{t}_\mu\left\{\partial^\lambda\partial^\sigma
\left[p_\nu\beta^\nu(x)-\zeta(x)\right]\right\} (\slashed{p}_2+m)\gamma^\mu(\slashed{p}_1+m) 
n_F(x,p_1)\left[1-n_F(x,p_2)\right]\int_0^1 \di z \; \e^{z\beta\cdot q}\,,\nonumber\\
f_D^\sigma(p_2,p_1)&\equiv&\frac{1}{4}\hat{t}_\mu\epsilon^{\mu\nu\lambda\rho}
\left[\partial^\sigma\Omega_{\nu\lambda}(x)\right] (\slashed{p}_2+m)\gamma_\rho\gamma^5 
(\slashed{p}_1+m) n_F(x,p_1)\left[1-n_F(x,p_2)\right]\int_0^1 \di z \; \e^{z\beta\cdot q}\,.
\end{eqnarray}
Integrating by parts, we can carry out the integral over $\di^4q$ in eq. (\ref{W-second-order}) 
and obtain:
\begin{eqnarray} \label{W-second-order-new}
W_\text{lin}^{(2)}(x,p)&=&-\frac{\delta(p^2-m^2)}{(2\pi)^3 2|p^0|}\lim_{q^\mu\rightarrow0}
\left[D_\lambda^q D_\sigma^qf_C^{\lambda\sigma}(p_2,p_1)-D_\sigma^q f_D^\sigma(p_2,p_1)\right] \nonumber\\
&&+\frac{\delta^{\prime}(p^2-m^2)}{(2\pi)^3 4|p^0|}\left[g_{\lambda\sigma}-\frac{2p_\lambda}
{p^0}g_{\sigma 0}+\frac{p_\lambda p_\sigma}{(p^0)^2}\right]\lim_{q^\mu\rightarrow0}
f_C^{\lambda\sigma}(p_2,p_1)\,,
\end{eqnarray}
where the operator $D^q$ is defined in eq. (\ref{operator-D}).  
We note that at the second order in gradient, the Wigner function features a term proportional to 
$\delta^{\prime}(p^2-m^2)$, that is it includes an off-shell contribution. 
This is not surprising because the four-momentum argument in the Wigner function is not 
necessarily on-shell. Nevertheless, as discussed below equation \eqref{Wigner-definition}, after 
its integration over an {\em arbitrary} hypersurface like in equation \eqref{spindef}, (i.e. 
$\int_\Sigma \di \Sigma \cdot p \; W(x,p)$) four-momentum $p$ becomes on-shell, so that the integrated 
off-shell part in the equation \eqref{W-second-order-new} which is proportional to $\delta^{\prime}(p^2-m^2)$ 
must vanish. Indeed, by choosing as integration hypersurface the hyperplane $t=\rm const$ one can 
show that if $\lambda=1,2,3$ or $\sigma=1,2,3$, its off-shell part after integration yields a 
boundary term which vanishes if the spatial region is sufficiently large. On the other hand, if 
$\lambda=\sigma=0$, the off-shell part vanishes altogether. Furthermore, in this case, by using
the explicit expression of $f_C^{\lambda\sigma}$ in (\ref{functions_f3_f4}), it can be shown that the 
term proportional to $\delta^{\prime}(p^2-m^2)$ only contributes to the scalar and vector components 
of the Wigner function, but not to the axial-vector one, which is relevant for the spin vector calculation.

Altogether, by using eq. \eqref{traceW0} and eq. \eqref{W-second-order-new} in the last equation 
of eq. (\ref{Spin-polarization}), after working out all traces one obtains:
\begin{eqnarray}\label{S-lin-second}
S_\text{lin}^{(2)\mu}(p)&=&\frac{1}{4m(p \cdot \hat t)^2N}\int \di \Sigma\cdot p_+\, n_F(x,p)
\left[1-n_F(x,p)\right] (y_\Sigma(0)-x)\cdot \hat t \nonumber\\
&&\times \hat{t}_\alpha p_\rho\left[\epsilon^{\mu\sigma\alpha\rho}p^\lambda p^\nu
\partial_\sigma\xi_{\nu\lambda}+\left(\frac{1}{2}p^\alpha\epsilon^{\mu\nu\lambda\rho}-\epsilon^{\mu\alpha\lambda\rho}p^\nu\right)p^\sigma\partial_\sigma\varpi_{\nu\lambda}\right. \nonumber\\
&&\left.-\epsilon^{\mu\sigma\alpha\rho}p^\lambda\partial_\sigma\partial_\lambda\zeta+\frac{1}{2}\epsilon^{\alpha\nu\lambda\sigma}\partial^\rho(\Omega_{\nu\lambda}-\varpi_{\nu\lambda})
(p^\mu p_\sigma-m^2g^\mu_\sigma)\right]\,,
\end{eqnarray}
where $N$ is shown in eq. \eqref{norm}.
The most striking feature of eq. \eqref{S-lin-second} is its explicit dependence 
on the average time $y^0_\Sigma(0)=y_\Sigma(0) \cdot \hat t$ of the space-like part of the 
hypersurface $\Sigma$. This finding, surprising as it seems, is indeed in full agreement with the 
expectation that a local equilibrium term, which is not supposed to survive at global equilibrium, 
shows a dependence on the shape of the hypersurface, as it was discussed at the end of
the previous subsection. It now appears that eq. \eqref{S-lin-second} is a pure relativistic 
effect: indeed, if $\Sigma$ is a flat 3D hypersurface, $x^0=y_\Sigma^0$ for any point $x$ 
on $\Sigma$, resulting in a vanishing $S_\text{lin}^{(2)\mu}$. The feature that $S^{(2)\mu}_\text{lin}(p)$ 
is proportional to $y^0_\Sigma(0)-x^0$ can also be understood as a consequence of space and time 
reversal symmetry (P and T symmetry). Since  $S^\mu(p,x^0)\rightarrow -S^\mu(p,-x^0)$ (where, 
for clarity, the dependence on $x^0$ is made explicit) under PT transformation, it must be constructed 
using PT-odd building blocks. At second order in gradients, these are 
$\partial_\mu\partial_\nu\beta_\rho, \partial_\mu\partial_\nu\zeta$, and $\partial_\mu\Omega_{\nu\rho}$
(which are all PT-even) multiplied by arbitrary powers of $p$ and odd powers of $y_\Sigma(0)-x$ 
(such combination is due to time translation symmetry). However, the power of $y_\Sigma(0)-x$ must be smaller than the order in gradients, implying that only the first power of $y_\Sigma(0)-x$ 
can appear in $S^{(2)\mu}_\text{lin}(p)$. Furthermore, the request of like transformations under 
P and T separately of both sides of equation \eqref{S-lin-second}, implies that only the time component $y^0_\Sigma(0)-x^0$ can appear. 

We now come to the spin polarization induced by quadratic terms in the gradient expansion.
According to eq. \eqref{Spin-polarization}, this reads: 
\begin{equation} \label{eq-Spin-polarization-quad}
S^{(2)\mu}_\text{quad}(p)=\frac{1}{2}\frac{\int \di\Sigma\cdot p_+\,
\tr\left[\gamma^\mu \gamma^5 W_\text{quad}(x,p)\right]}{\int \di\Sigma\cdot p_+\,
\tr[W^{(0)}(x,p)]}-S^{(1)\mu}(p)\frac{\int \di\Sigma\cdot p_+\,\tr[W^{(1)}(x,p)]}
{\int \di\Sigma\cdot p_+\,\tr[W^{(0)}(x,p)]}\,,
\end{equation}
where the second term arises because also the particle number is modified at the first 
order in the gradient expansion. The second term on the right hand side of eq.
\eqref{eq-Spin-polarization-quad} which can be obtained from eqs. \eqref{traceW0}, \eqref{Wigner-components-1},
and \eqref{eq:first-order-polarization} does not need 
further explanation. On the other hand, the function $W_\text{quad}$ is obtained by 
plugging $\mathcal{B}^{(1)}$ in eq. (\ref{eq:B-first-order}) into eq. (\ref{eq:W-quad-final}). 
The expression of $W_\text{quad}$ is too complicated to be listed here, however we 
have found a compact formula for the axial-vector component of $W_\text{quad}$ as a 
function of already known quantities, whose derivation is reported in the Appendix 
\ref{app:quad}:
\begin{equation}\label{Wigner-quad-2}
\tr\left[\gamma^\mu \gamma^5 W_\text{quad}(x,p)\right]=\frac{\left[1-2n_F(x,p)\right]
\tr\left[\gamma^\mu \gamma^5 W^{(1)}(x,p)\right]\tr\left[W^{(1)}(x,p)\right]}
{\left[1-n_F(x,p)\right]\tr\left[W^{(0)}(x,p)\right]}\,.
\end{equation}
Plugging the above expression into $S^{(2)\mu}_\text{quad}$ in eq. (\ref{eq-Spin-polarization-quad}), 
one can obtain the spin polarization induced by quadratic terms of first order quantities. 
Note that, according to eq. \eqref{Wigner-components-1}, if $\Sigma$ is a flat 3D hypersurface 
with normal vector $\hat{t}^\mu=(1,0,0,0)$, $\tr\left[W^{(1)}(x,p)\right]=0$ because of 
vanishing $y^0_\Sigma(0)-x^0$ for any point $x$ on $\Sigma$, leading to vanishing $S_\text{quad}$. 
Therefore, for such a flat hypersurface, the total spin polarization at second order, given by the sum $S^{(2)\mu}=S^{(2)\mu}_\text{lin}+S^{(2)\mu}_\text{quad}$, vanishes. This is in agreement
with calculations at the global equilibrium, which are notably independent of the hypersurface,
which show that there is no quadratic contribution in thermal vorticity for the spin polarization
vector \cite{Palermo:2023cup}.

\section{Summary} \label{sec:Summary}

In summary, we have derived an expression of the average spin polarization vector of spin-1/2 
fermions at local thermodynamic equilibrium up to the second order in the gradient expansion of
hydro-thermodynamic fields, upgrading previous results at the leading order. We have calculated
both the contributions from the quadratic term in the gradients and the linear in the second 
order derivatives. 

The calculations in this work are based on the local thermodynamic equilibrium density operator
and therefore have an expected dependence on the freeze-out hypersurface $\Sigma$. 
Under some geometric approximation, we find that the second order correction $S^{(2)\mu}$ depends 
on the time coordinate $x^0$ of the centre-of-mass frame for any point $x$ on $\Sigma$. Notably, for 
the case of $\Sigma$ coinciding with a hyperplane with a constant normal vector $\hat{t}^\mu=(1,0,0,0)$, 
the term $S^{(2)\mu}$ vanishes. Nevertheless, in relativistic heavy ion collision $\Sigma$ is neither 
flat nor purely space-like, resulting in a non-vanishing $S^{(2)\mu}$. 

In relativistic heavy ion collisions, while the quadratic term is expected to be a small correction 
to the leading order in view of the smallness of $\varpi$ and $\xi$ at the freeze-out, as shown 
by numerical simulations, the impact of the second order derivative term is not easy to estimate 
and requires numerical computations.

\section*{Acknowledgments}

X.L.S. and F. B. are partly supported by ICSC - {\it Centro Nazionale di Ricerca in High Performance Computing, Big Data and Quantum Computing}, funded by European Union - NextGenerationEU and by the project PRIN2022 {\it Advanced Probes of the Quark Gluon Plasma} funded by ``Ministero dell'Universit\`a e della Ricerca". Z.-H.Z. and X.-G.H. are supported by the National Key Research and Development Program of China (Grant No. 2022YFA1604900), the Natural Science Foundation of Shanghai (Grant No. 23JC1400200), and the National Natural Science Foundation of China (Grant No. 12147101, No. 12225502, and No. 12075061).

\bibliographystyle{apsrev}
\bibliography{ref}

\appendix

\section{Integral over the hypersurface}
\label{app:hyp}

\begin{figure}
  \includegraphics[width=0.5\textwidth]{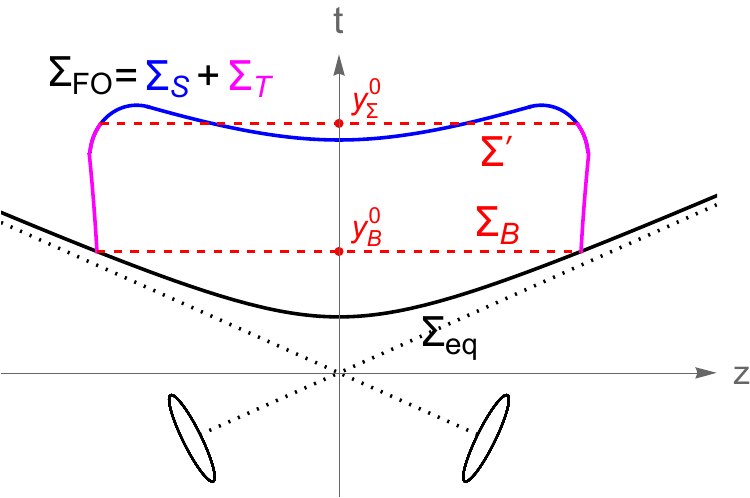}
  \caption{\label{fig:freezeout2} Spacetime diagram of a relativistic heavy-ion collision. Here $\Sigma_\text{eq}$ (black line) 
  is the hypersurface where LTE is achieved and $\Sigma_\text{FO}$ is the hypersurface where the quark-gluon plasma decouples, which consists of the space-like part $\Sigma_S$ (blue line) and the time-like parts $\Sigma_T$ (magenta lines). For calculating the integral over $\Sigma_\text{FO}$, we introduce two flat 
  hypersurfaces $\Sigma_B$ and $\Sigma^\prime$ (red dashed lines) with constant times $y_B^0$ and $y_\Sigma^0$, respectively.}
\end{figure}

The aim of this section is to calculate the integral over the freeze-out hypersurface:
\begin{equation}
\int_{\Sigma} \di \Sigma_{\mu}\; \e^{-\ii q\cdot(y-x)}\,.\label{eq:integral}
\end{equation}
The hypersurface $\Sigma$ can be split into a space-like part $\Sigma_{S}$ and a time-like 
part $\Sigma_{T}$ (there is also a light-like part, but it is 2D surface with vanishing 
measure). It is convenient to introduce two auxiliary hypersurfaces $\Sigma_{B}$ and 
$\Sigma^{\prime}$, which are portions of hyperplanes with constant times $y_{B}^{0}$ 
and $y_{\Sigma}^{0}$ in the centre-of-mass frame, see fig. \ref{fig:freezeout2}. The hypersurface 
$\Sigma_B$ is defined as the portion of hyperplane joining the intersections between the initial local 
equilibrium hypersurface and the freeze-out, while $\Sigma^\prime$ is a portion of  
hyperplane, with $\hat t$ as normal vector, which intersects the hypersurface $\Sigma$ at
some 2D boundary. Obviously, for any point $y^{\mu}$ on $\Sigma^{\prime}$, the time component 
$y^{0}$ is a constant. The idea is that, in general, it is possible to approximate:
\begin{equation}\label{matching}
\int_{\Sigma^{\prime}} \di \Sigma_{\mu}\; \e^{-\ii q\cdot(y-x)} \simeq
\int_{\Sigma_{S}} \di \Sigma_{\mu}\; \e^{-\ii q\cdot(y-x)}\,.
\end{equation}
and that $y^0$ can be chosen on $\Sigma^\prime$ so as to make this approximation an equality
for the time component of the equation \eqref{matching}. Any point on $\Sigma_{S}$ has a time 
component which is a function of the space Cartesian coordinates ${\bf y}$, that is 
$y^{0}=y^{0}({\bf y})$. Thus, the $\mu=0$ component of eq. \eqref{matching} can be rewritten as:
\be\label{matching2}
\int \di^{3}{\rm y}\; \e^{-\ii q^{0} (y_{\Sigma}^{0}-x^{0})}
\e^{\ii{\bf q}\cdot({\bf y}-{\bf x})}
= \int \di^{3}{\rm y}\; \e^{-\ii q^{0}\left( y^{0}({\bf y})- x^0 \right)} 
\e^{\ii{\bf q}\cdot({\bf y}-{\bf x})}
\ee
with unknown $y_\Sigma^0$. In the equation \eqref{matching2} it can be shown that, by 
using $y^i, \; i=1,2,3$ as variables to describe the hypersurface $\Sigma_S$, it turns out 
that $\di \Sigma_0 = \di^3 y$ without any other multiplicative factor. In general, the 
integration domains in ${\bf y}$ can be different, yet this equation should have a solution 
for each $q$, that is $y_\Sigma^0(q)$. In the limit of $q^{\mu}\rightarrow 0$, and assuming
that the domains of $\bf y$ are approximately the same, one can rewrite eq. \eqref{matching2} as:
$$
\int \di^{3}{\rm y}\; \left( \e^{-\ii q^{0} y_{\Sigma}^{0}} 
- \e^{-\ii q^{0} y^{0}({\bf y})} \right) \simeq q^0 \int \di^{3}{\rm y}\; 
 \left( y_{\Sigma}^{0} - y^{0}({\bf y}) \right) = 0
$$
giving as average time for $\Sigma_{S}$:
$$
y_{\Sigma}^{0}(0) \simeq \frac{\int_{\Sigma_{S}}\di \Sigma \; y^{0}({\bf y})}
{\int_{\Sigma_{S}}\di \Sigma}\,.
$$

By using eq. \eqref{matching}, the integral in eq. (\ref{eq:integral}) can be
now cast in the following form:
$$
\int_{\Sigma} \di \Sigma_{\mu}\; \e^{-\ii q\cdot(y-x)} \simeq
\int_{\Sigma^\prime \cup \Sigma_L} \di \Sigma_{\mu}\; \e^{-iq\cdot(y-x)}\,,
$$
where $\Sigma_L$ is the part of $\Sigma$ joining $\Sigma^\prime$ and $\Sigma_B$
(which is approximate to $\Sigma_T$, see fig. \ref{fig:freezeout2}). Using the Gauss' theorem, one can convert the above
3D integration to a 3D integration over $\Sigma_B$ plus a 4D integral over the region 
encompassed by $\Sigma^{\prime}$, $\Sigma_{L}$, and $\Sigma_{B}$. Furthermore, if the hypersurface 
$\Sigma_{B}$ is large enough, we can approximate the integrals with their infinite
space extension limits and obtain:
\begin{align*}
\int_{\Sigma}\di \Sigma_{\mu}\; \e^{-iq\cdot(y-x)} & \simeq \int_{\Sigma^\prime \cup \Sigma_L} 
\di \Sigma_{\mu}\; \e^{-iq\cdot(y-x)} \simeq \hat{t}_{\mu} \e^{-\ii q^{0}(y_{B}^{0}-x^{0})}\int \di^{3}{\rm y}\; 
\e^{\ii{\bf q}\cdot({\bf y}-{\bf x})} +\int_{y_{B}^{0}}^{y_{\Sigma}^{0}(q)} \!\! 
\di y^{0} \int \di^{3}{\rm y}\; \frac{\partial}{\partial y^\mu} \e^{-\ii q\cdot(y-x)}\,
\\
& \simeq (2\pi)^3\left[\hat{t}_\mu \e^{-\ii q^0(y_B^0-x^0)}
-\ii q_\mu \int_{y_B^0}^{y_\Sigma^0(q)}\di y^0 \; \e^{-\ii q^0(y^0-x^0)}\right] \delta^{3}({\bf q})
= (2\pi)^3 \e^{-\ii q^0(y_\Sigma^0-x^0)}\delta^{3}({\bf q})\hat{t}_\mu\,,
\end{align*}
which is eq. (\ref{eq:hypersurface-int}) in the main text.

\section{Calculation of $W_{\text{lin}}^{(2)}$}
\label{app:w2lin}

In this section we show some of most important steps of the evaluation of $W_{\text{lin}}^{(2)}$ 
in eq. (\ref{W-second-order}). We first focus on the contribution from $f_{D}^{\sigma}$, which 
is given by
$$
I_{D} = \frac{1}{(2\pi)^{3}}\int \di^{4}q \; \delta(p_{1}^{2}-m^{2})\delta(p_{2}^{2}-m^{2})\, 
\text{sgn}(p_{1}^{0}) \, \text{sgn}(p_{2}^{0}) f_{D}^{\sigma}(p_{2},p_{1})i\partial_{\sigma}^{q}
\left[\e^{-iq^{0}(y_{\Sigma}^{0}-x^{0})}\delta^{3}({\bf q})\right]\,.
$$
Integrating by parts, the above formula can be turned into:
\begin{eqnarray}\label{IDone}
I_{D} & = & -\frac{1}{(2\pi)^{3}}\int \di^{4}q \; \e^{-\ii q^{0}(y_{\Sigma}^{0}-x^{0})}
\delta^{3}({\bf q})\text{sgn}(p_{1}^{0})\text{sgn}(p_{2}^{0}) \left\{ 
\left[i\partial_{\sigma}^{q}f_{D}^{\sigma}(p_{2},p_{1})\right]\delta(p_{1}^{2}-m^{2})
\delta(p_{2}^{2}-m^{2})\right.\nonumber \\
 &  & \left.+f_{D}^{\sigma}(p_{2},p_{1})\ii \partial_{\sigma}^{q}\left[ \delta(p_{1}^{2}-m^{2})
 \delta(p_{2}^{2}-m^{2})\right]\right\} \,.
\end{eqnarray}
In this transformation, we have neglected derivatives of the sign functions, because they 
yield delta functions of an energy equation that in general cannot be fulfilled due to the on-shell conditions enforced by other delta functions. Rewriting 
the product of two delta functions as in eq. \eqref{deltaprod}, we can calculate 
their momentum derivative as follows:
\begin{eqnarray*}
&& \ii \partial_{\sigma}^{q}\left[ \delta(p_{1}^{2}-m^{2})\delta(p_{2}^{2}-m^{2}) \right]= 
\ii \frac{q_{\sigma}}{4|p^{0}|}\delta^{\prime}\left(p^{2}+\frac{q^{2}}{4}-m^{2}\right)
\delta\left(q^{0}-\frac{{\bf p}\cdot{\bf q}}{p^{0}}\right)+\frac{ip_{\sigma}}{2|p^{0}|p^{0}}
\delta\left(p^{2}+\frac{q^{2}}{4}-m^{2}\right)\delta^{\prime}\left(q^{0}-\frac{{\bf p}\cdot{\bf q}}
{p^{0}}\right) \nonumber \\
 && = \frac{p_{\sigma}}{p^{0}}i\partial_{0}^{q}\left[\frac{1}{2|p^{0}|}
 \delta\left(p^{2}+\frac{q^{2}}{2}-m^{2}\right)\delta\left(q^{0}-\frac{{\bf p}\cdot{\bf q}}{p^{0}}\right)\right]
 +i\frac{q_{\sigma}}{4|p^{0}|}\delta^{\prime}\left(p^{2}+\frac{q^{2}}{4}-m^{2}\right)
 \delta\left(q^{0}-\frac{{\bf p}\cdot{\bf q}}{p^{0}}\right)\nonumber \\
 && -\frac{ip_{\sigma}q^{0}}{4|p^{0}|p^{0}}\delta^{\prime}\left(p^{2}+\frac{q^{2}}{4}-m^{2}\right)
 \delta\left(q^{0}-\frac{{\bf p}\cdot{\bf q}}{p^{0}}\right)\,.
\end{eqnarray*}
When plugging the above formula back into $I_{D}$, we find that $\delta\left(q^{0}-{\bf p}
\cdot{\bf q}/p^{0}\right)$ and $\delta({\bf q})$ involve the restriction $q^{\mu}=0$, implying 
that the last two terms on the right-hand side do not contribute to $I_{D}$. Therefore,
the \eqref{IDone} becomes:
\begin{eqnarray}\label{IDtwo}
I_{D} & = & -\frac{1}{(2\pi)^{3}}\int \di^{4}q \; \e^{-iq^{0}(y_{\Sigma}^{0}-x^{0})}\delta^{3}({\bf q})
\text{sgn}(p_{1}^{0})\text{sgn}(p_{2}^{0})
\left\{ \left[i\partial_{\sigma}^{q}f_{D}^{\sigma}(p_{2},p_{1})\right]\frac{1}{2|p^{0}|}
 \delta\left(p^{2}+\frac{q^{2}}{2}-m^{2}\right)\delta\left(q^{0}-\frac{{\bf p}\cdot{\bf q}}{p^{0}}
 \right)\right.\nonumber \\
 &  & \left.+f_{D}^{\sigma}(p_{2},p_{1})\frac{p_{\sigma}}{p^{0}}i\partial_{0}^{q}\left[\frac{1}{2|p^{0}|}
 \delta\left(p^{2}+\frac{q^{2}}{2}-m^{2}\right)\delta\left(q^{0}-\frac{{\bf p}\cdot{\bf q}}{p^{0}}\right)
 \right]\right\} \,.
\end{eqnarray}
For the derivative with respect to $q^{0}$ in the \eqref{IDtwo}, we integrate by parts again:
\begin{eqnarray*}
I_{D} & = & -\frac{1}{(2\pi)^{3}}\int \di^4 q \; \e^{-iq^{0}(y_{\Sigma}^{0}-x^{0})}\delta^{3}({\bf q})
\text{sgn}(p_{1}^{0})\text{sgn}(p_{2}^{0}) \frac{1}{2|p^{0}|}\delta\left(p^{2}+\frac{q^{2}}{2}-m^{2}\right)
\delta\left(q^{0}-\frac{{\bf p}\cdot{\bf q}}{p^{0}}\right)\nonumber \\
 &  & \times\left[i\partial_{\sigma}^{q}-\frac{p_{\sigma}}{p^{0}}(i\partial_{0}^{q}+y_{\Sigma}^{0}-x^{0})\right]
 f_{D}^{\sigma}(p_{2},p_{1})\,.
\end{eqnarray*}
to finally get:
\begin{equation}\label{IDthree}
I_{D}=\frac{\delta(p^{2}-m^{2})}{(2\pi)^{3}2|p^{0}|}\lim_{q^{\mu}\rightarrow0}D_{\sigma}^{q}
f_{D}^{\sigma}(p_{2},p_{1})\,,
\end{equation}
where the operator $D_{\sigma}^{q}$ is defined in eq. (\ref{operator-D}).

We now turn to the contribution from $f_{C}^{\lambda\sigma}$, which is given by:
$$
I_{C} = \frac{1}{(2\pi)^{3}}\int \di^{4}q \; 
\delta(p_{1}^{2}-m^{2})\delta(p_{2}^{2}-m^{2})\text{sgn}(p_{1}^{0})\text{sgn}(p_{2}^{0})
f_{C}^{\lambda\sigma}(p_{2},p_{1})\partial_{\lambda}^{q}\partial_{\sigma}^{q}
\left[\e^{-\ii q^{0}(y_{\Sigma}^{0}-x^{0})}\delta^{3}({\bf q})\right]\,.
$$
Similarly for the foregoing case, the above integral can be carried out by parts 
to obtain:
\begin{eqnarray*}
I_{C} & = & \frac{1}{(2\pi)^{3}}\int \di^{4}q \; \e^{-\ii q^{0}(y_{\Sigma}^{0}-x^{0})}
\delta^{3}({\bf q}) \text{sgn}(p_{1}^{0})\text{sgn}(p_{2}^{0}) 
\left\{\left[\partial_{\lambda}^{q}\partial_{\sigma}^{q}f_{C}^{\lambda\sigma}(p_{2},p_{1})\right]
\delta(p_{1}^{2}-m^{2})\delta(p_{2}^{2}-m^{2}) \right. \nonumber \\
&& \left. +2\left[\partial_{\lambda}^{q}f_{C}^{\lambda\sigma}(p_{2},p_{1})\right]
\left[\partial_{\sigma}^{q}\delta(p_{1}^{2}-m^{2})\delta(p_{2}^{2}-m^{2})\right]
+f_{C}^{\lambda\sigma}(p_{2},p_{1})\left[\partial_{\lambda}^{q}\partial_{\sigma}^{q}
\delta(p_{1}^{2}-m^{2})\delta(p_{2}^{2}-m^{2})\right] \right\} \,,
\end{eqnarray*}
where the momentum derivatives of sign functions have been neglected for the same
reason as for the previous case, and the symmetry $f_{C}^{\lambda\sigma}=f_{C}^{\sigma\lambda}$
has been used. The first and second terms on the right-hand-side of the above formula 
can be simplified in a similar way as for the previous calculation of $I_{D}$. 
Then $I_{C}$ is converted to:
\begin{eqnarray}\label{eq:IC}
I_{C} & = & \frac{\delta(p^{2}-m^{2})}{(2\pi)^{3}2|p^{0}|}\lim_{q^{\mu}\rightarrow0}
\left[\partial_{\lambda}^{q}\partial_{\sigma}^{q} f_{C}^{\lambda\sigma}(p_{2},p_{1})+
2i\frac{p_{\sigma}}{p^{0}}(i\partial_{0}^{q}+y_{\Sigma}^{0}-x^{0})\partial_{\lambda}^{q}
f_{C}^{\lambda\sigma}(p_{2},p_{1})\right]\nonumber \\
 &  & +\frac{1}{(2\pi)^{3}}\int \di^{4}q \; \e^{-\ii q^{0}(y_{\Sigma}^{0}-x^{0})}\delta^{3}({\bf q})
 \text{sgn}(p_{1}^{0})\text{sgn}(p_{2}^{0}) f_{C}^{\lambda\sigma}(p_{2},p_{1})
 \left[\partial_{\lambda}^{q}\partial_{\sigma}^{q}\delta(p_{1}^{2}-m^{2})
 \delta(p_{2}^{2}-m^{2})\right]\,.
\end{eqnarray}
For the last term on the right hand side, we have:
\begin{eqnarray}\label{eq:derivative-1}
&& \partial_{\lambda}^{q}\partial_{\sigma}^{q}\delta(p_{1}^{2}-m^{2})\delta(p_{2}^{2}-m^{2}) 
 = \frac{q_{\lambda}q_{\sigma}}{8|p^{0}|}\delta^{\prime\prime}\left(p^{2}+
\frac{q^{2}}{4}-m^{2}\right)\delta\left(q^{0}-\frac{{\bf p}\cdot{\bf q}}{p^{0}}\right)+
\frac{g_{\lambda\sigma}}{4|p^{0}|}\delta^{\prime}\left(p^{2}+\frac{q^{2}}{4}-m^{2}\right)
\delta\left(q^{0}-\frac{{\bf p}\cdot{\bf q}}{p^{0}}\right)\nonumber \\
&& +\frac{p_{\lambda}p_{\sigma}}{2(p^{0})^{2}|p^{0}|}\delta\left(p^{2}+\frac{q^{2}}{4}-m^{2}\right)
\delta^{\prime\prime}\left(q^{0}-\frac{{\bf p}\cdot{\bf q}}{p^{0}}\right)
+\frac{q_{\lambda}p_{\sigma}+p_{\lambda}q_{\sigma}}{4p^{0}|p^{0}|}
\delta^{\prime}\left(p^{2}+\frac{q^{2}}{4}-m^{2}\right)
\delta^{\prime}\left(q^{0}-\frac{{\bf p}\cdot{\bf q}}{p^{0}}\right)\,.
\end{eqnarray}
In turn, the third and last terms on the right hand side of eq.~\eqref{eq:derivative-1} 
can be worked out as follows:
\begin{eqnarray}\label{eq:derivative-2}
 && \frac{p_{\lambda}p_{\sigma}}{2(p^{0})^{2}|p^{0}|}
\delta\left(p^{2}+\frac{q^{2}}{4}-m^{2}\right)\delta^{\prime\prime}
\left(q^{0}-\frac{{\bf p}\cdot{\bf q}}{p^{0}}\right)
 = \partial_{q}^{0}\partial_{q}^{0}\left[\frac{p_{\lambda}p_{\sigma}}{2(p^{0})^{2}|p^{0}|}
 \delta\left(p^{2}+\frac{q^{2}}{4}-m^{2}\right)\delta\left(q^{0}-\frac{{\bf p}
 \cdot{\bf q}}{p^{0}}\right)\right]\nonumber \\
 & & -\partial_{q}^{0}\left[\frac{q^{0}p_{\lambda}p_{\sigma}}{2(p^{0})^{2}|p^{0}|}
 \delta^{\prime}\left(p^{2}+\frac{q^{2}}{4}-m^{2}\right)\delta\left(q^{0}-
 \frac{{\bf p}\cdot{\bf q}}{p^{0}}\right)\right]
 +\frac{(q^{0})^{2}p_{\lambda}p_{\sigma}}{8(p^{0})^{2}|p^{0}|}\delta^{\prime\prime}
 \left(p^{2}+\frac{q^{2}}{4}-m^{2}\right)\delta\left(q^{0}-\frac{{\bf p}\cdot{\bf q}}{p^{0}}\right)
 \nonumber \\
 && +\frac{p_{\lambda}p_{\sigma}}{4(p^{0})^{2}|p^{0}|}\delta^{\prime}
 \left(p^{2}+\frac{q^{2}}{4}-m^{2}\right)\delta\left(q^{0}-\frac{{\bf p}\cdot{\bf q}}{p^{0}}\right)\,,
\end{eqnarray}
and
\begin{eqnarray}\label{eq:derivative-3}
 && \frac{q_{\lambda}p_{\sigma}+p_{\lambda}q_{\sigma}}{4p^{0}|p^{0}|}\delta^{\prime}
 \left(p^{2}+\frac{q^{2}}{4}-m^{2}\right)\delta^{\prime}\left(q^{0}-\frac{{\bf p}\cdot{\bf q}}{p^{0}}\right)
 = \partial_{q}^{0}\left[\frac{q_{\lambda}p_{\sigma}+p_{\lambda}q_{\sigma}}{4p^{0}|p^{0}|}
 \delta^{\prime}\left(p^{2}+\frac{q^{2}}{4}-m^{2}\right)
 \delta\left(q^{0}-\frac{{\bf p}\cdot{\bf q}}{p^{0}}\right)\right]\nonumber \\
 && -\frac{g_{\lambda0}p_{\sigma}+p_{\lambda}g_{\sigma0}}{4p^{0}|p^{0}|}\delta^{\prime}
 \left(p^{2}+\frac{q^{2}}{4}-m^{2}\right)\delta\left(q^{0}-\frac{{\bf p}\cdot{\bf q}}{p^{0}}\right)
 -\frac{q^{0}(q_{\lambda}p_{\sigma}+p_{\lambda}q_{\sigma})}{8p^{0}|p^{0}|}\delta^{\prime\prime}
 \left(p^{2}+\frac{q^{2}}{4}-m^{2}\right)\delta\left(q^{0}-\frac{{\bf p}\cdot{\bf q}}{p^{0}}\right)\,.
\end{eqnarray}
Subsituting eqs. (\ref{eq:derivative-1})-(\ref{eq:derivative-3}) into eq. (\ref{eq:IC}) and 
using the fact that $\delta\left(q^{0}-{\bf p}\cdot{\bf q}/p^{0}\right)$ and $\delta^{3}({\bf q})$ 
restricts $q^{\mu}=0$, we arrive at:
\begin{eqnarray*}
&& I_{C}=\frac{\delta(p^{2}-m^{2})}{(2\pi)^{3}2|p^{0}|}\lim_{q^{\mu}\rightarrow0}
\left[\partial_{\lambda}^{q}\partial_{\sigma}^{q} f_{C}^{\lambda\sigma}(p_{2},p_{1})+
2i\frac{p_{\sigma}}{p^{0}}(i\partial_{0}^{q}+y_{\Sigma}^{0}-x^{0})\partial_{\lambda}^{q}
f_{C}^{\lambda\sigma}(p_{2},p_{1})\right]\nonumber \\
&& +\frac{1}{(2\pi)^{3}}\int \di^{4}q \; \e^{-iq^{0}(y_{\Sigma}^{0}-x^{0})}\delta^{3}({\bf q})
\text{sgn}(p_{1}^{0})\text{sgn}(p_{2}^{0})f_{C}^{\lambda\sigma}(p_{2},p_{1})
\left\{ \partial_{q}^{0}\partial_{q}^{0}\left[\frac{p_{\lambda}p_{\sigma}}{2(p^{0})^{2}|p^{0}|}
\delta\left(p^{2}+\frac{q^{2}}{4}-m^{2}\right)\delta\left(q^{0}-\frac{{\bf p}\cdot{\bf q}}{p^{0}}
\right)\right]\right.\nonumber \\
&& \left.+\frac{1}{4|p^{0}|}\left[g_{\lambda\sigma}-\frac{g_{\lambda0}p_{\sigma}+
p_{\lambda}g_{\sigma0}}{p^{0}}+\frac{p_{\lambda}p_{\sigma}}{(p^{0})^{2}}\right]\delta^{\prime}\left(p^{2}+
\frac{q^{2}}{4}-m^{2}\right)\delta\left(q^{0}-\frac{{\bf p}\cdot{\bf q}}{p^{0}}\right)\right\} \,.
\end{eqnarray*}
Finally,by integrating by parts again and carrying out the integral over $\di^{4}q$, we 
obtain
\begin{align*}
I_{C} =& \frac{\delta(p^{2}-m^{2})}{(2\pi)^{3}2|p^{0}|}\lim_{q^{\mu}\rightarrow0}
\left[\partial_{\lambda}^{q}\partial_{\sigma}^{q} +2i\frac{p_{\sigma}}{p^{0}}
(i\partial_{0}^{q}+y_{\Sigma}^{0}-x^{0})\partial_{\lambda}^{q}-\frac{p_{\lambda}p_{\sigma}}
{(p^{0})^{2}}(i\partial_{0}^{q}+y_{\Sigma}^{0}-x^{0})^{2}\right]
f_{C}^{\lambda\sigma}(p_{2},p_{1}) \\ \nonumber
&+\frac{\delta^{\prime}(p^{2}-m^{2})}{(2\pi)^{3}4|p^{0}|}
\lim_{q^{\mu}\rightarrow0}\left[g_{\lambda\sigma}-\frac{g_{\lambda0}p_{\sigma}+p_{\lambda}
g_{\sigma0}}{p^{0}}+\frac{p_{\lambda}p_{\sigma}}{(p^{0})^{2}}\right] f_{C}^{\lambda\sigma}(p_{2},p_{1})\,.
\end{align*}
which can be expressed in a compact form using the operator $D_{\sigma}^{q}$ in eq. (39),
\be\label{ICtwo}
I_{C}  = -\frac{\delta(p^{2}-m^{2})}{(2\pi)^{3}2|p^{0}|}\lim_{q^{\mu}\rightarrow0}
D_{\lambda}^{q}D_{\sigma}^{q}f_{C}^{\lambda\sigma}(p_{2},p_{1})
+\frac{\delta^{\prime}(p^{2}-m^{2})}{(2\pi)^{3}4|p^{0}|}\lim_{q^{\mu}\rightarrow0}
\left[g_{\lambda\sigma}-\frac{2p_{\lambda}}{p^{0}}g_{\sigma0}+
\frac{p_{\lambda}p_{\sigma}}{(p^{0})^{2}}\right]f_{C}^{\lambda\sigma}(p_{2},p_{1})\,.
\ee
The Wigner function $W_{\text{lin}}^{(2)}$ in (\ref{W-second-order-new}) is then obtained
by summing eqs. \eqref{ICtwo} and \eqref{IDthree}.

\section{Calculation of $W_{\text{quad}}$}
\label{app:quad}

To make notations more compact, we first define:
\begin{eqnarray*}
\Gamma_{1,\lambda}(p) & \equiv & \hat{t}_{\mu}\gamma^{\mu}[p^{\nu}
\partial_{\lambda}\beta_{\nu}(x)-\partial_{\lambda}\zeta(x)]\,,\nonumber \\
\Gamma_{2} & \equiv & \frac{1}{4}\hat{t}_{\mu}\epsilon^{\mu\nu\lambda\sigma}
\gamma_{\sigma}\gamma^{5}\Omega_{\nu\lambda}(x)\,,
\end{eqnarray*}
where the dependence of $\Gamma_{1,\lambda}$ and $\Gamma_2$ on $x$ is understood; 
also, $\Gamma_{1,\lambda}$ depends on $p$ whereas $\Gamma_{2}$ is independent of $p$. 
Thereby, the first order part $\mathcal{B}^{(1)}$ in eq. (\ref{eq:B-first-order}) can be rewritten 
as:
$$
\mathcal{B}^{(1)}(p_{2},p_{1})=-\left[\Gamma_{1,\lambda}(p) \ii 
\partial_{q}^{\lambda}+\Gamma_{2}\right]\left[\e^{-iq^{0}(y_{\Sigma}^{0}-x^{0})}
\delta^{3}({\bf q})\right]\,.
$$
Substituting it into $W_{\text{quad}}$ in eq. (\ref{eq:W-quad-final}), one obtains:
\begin{eqnarray}\label{eq:W-2nd-quad}
W_{\text{quad}}(x,p) & = & \int \di Q \; (\slashed{p}_{3}+m)
\left\{ \left[\Gamma_{1,\lambda}(p-q/2)i\partial_{q^{\prime}}^{\lambda}+\Gamma_{2}\right]
\left[\e^{-iq^{\prime0}(y_{\Sigma}^{0}-x^{0})}\delta^{3}({\bf q}^{\prime})\right]\right\} \nonumber \\
 &  & \times(\slashed{p}_{2}+m)\left\{ \left[\Gamma_{1,\sigma}(p+q^{\prime}/2)
 i\partial_{q}^{\sigma}+\Gamma_{2}\right]\left[\e^{-iq^{0}(y_{\Sigma}^{0}-x^{0})}
 \delta^{3}({\bf q})\right]\right\} (\slashed{p}_{1}+m)\,,
\end{eqnarray}
where $p_1\equiv p+(q+q^\prime)/2$, $p_2\equiv p+(-q+q^\prime)/2$, $p_3\equiv p-(q+q^\prime)/2$,
and
\begin{eqnarray*}
\di Q & \equiv & \frac{1}{2(2\pi)^{3}} \di^{4}q \di^{4}q^{\prime}
\delta(p_{1}^{2}-m^{2})\delta(p_{2}^{2}-m^{2})\delta(p_{3}^{2}-m^{2})\text{sgn}(p_{1}^{0})
\text{sgn}(p_{2}^{0}) \text{sgn}(p_{3}^{0})\nonumber \\
&& \times n_{F}(p)[1-n_{F}(p)][1-2n_{F}(p)]\left[1+\frac{1-4\e^{\beta\cdot p}+
\e^{2\beta\cdot p}}{6\left(\e^{2\beta\cdot p}-1\right)}\beta\cdot(q-q^{\prime})\right]\,.
\end{eqnarray*}
The $\di Q$ in the above equation has been obtained through the Taylor expansion 
in $q^{\mu}$ or $q^{\prime\mu}$ of the following expression:
\begin{eqnarray}\label{eq:expansion}
 &  & n_{F}(p_{1})[1-n_{F}(p_{3})]\int_{0}^{1}\di z_{1}\int_{0}^{z_{1}}\di z_{2}\ 
 \left\{ -\e^{z_{2}\beta\cdot q+z_{1}\beta\cdot q^{\prime}}n_{F}(p_{2})+\e^{z_{2}\beta\cdot q^{\prime}
 +z_{1}\beta\cdot q}[1-n_{F}(p_{2})]\right\} \nonumber \\
 & \simeq & \frac{1}{2}n_{F}(p)[1-n_{F}(p)][1-2n_{F}(p)]\left[1+\frac{1-4\e^{\beta\cdot p}+
 \e^{2\beta\cdot p}}{6\left(\e^{2\beta\cdot p}-1\right)}\beta\cdot(q-q^{\prime})+
 \mathcal{O}(q^{2},q^{\prime2},q\cdot q^{\prime})\right]\,,
\end{eqnarray}
where $\mathcal{O}(q^{2},q^{\prime2},q\cdot q^{\prime})$ terms can be dropped because 
they will not contribute to later calculations. In eq. (\ref{eq:W-2nd-quad}), 
$\Gamma_{1,\lambda}$ is multiplied
with a momentum derivative acting on $\delta^{3}({\bf q})$ or $\delta^{3}({\bf q}^{\prime})$,
while $\Gamma_{2}$ does not. The momentum derivative gives rise to complicated expressions when trying 
to integrate out the delta functions, thus we separate $W_{\text{quad}}$ into three parts:
\begin{equation*}
W_{\text{quad}}=W_{\text{quad},11}+W_{\text{quad},12}+W_{\text{quad},22}\,,
\end{equation*}
with:
\begin{eqnarray*}
&& W_{\text{quad,11}}(x,p) = \int \di Q \;(\slashed{p}_{3}+m)
\left\{ \left[\Gamma_{1,\lambda}(p-q/2)i\partial_{q^{\prime}}^{\lambda}\right]\left[\e^{-iq^{\prime0}
(y_{\Sigma}^{0}-x^{0})}\delta^{3}({\bf q}^{\prime})\right]\right\} \nonumber \\
 &  & \times(\slashed{p}_{2}+m)\left\{ \left[\Gamma_{1,\sigma}(p+q^{\prime}/2)i\partial_{q}^{\sigma}
 \right]\left[\e^{-iq^{0}(y_{\Sigma}^{0}-x^{0})}\delta^{3}({\bf q})\right]\right\} (\slashed{p}_{1}+m)\,,\\
&& W_{\text{quad,12}}(x,p) =  \int \di Q \;(\slashed{p}_{3}+m)\left\{ \left[\Gamma_{1,\lambda}(p-q/2)i\partial_{q^{\prime}}^{\lambda}\right]\left[\e^{-iq^{\prime0}(y_{\Sigma}^{0}-x^{0})}
\delta^{3}({\bf q}^{\prime})\right]\right\} (\slashed{p}_{2}+m)\Gamma_{2}(\slashed{p}_{1}+m)
\left[\e^{-iq^{0}(y_{\Sigma}^{0}-x^{0})}\delta^{3}({\bf q})\right]\nonumber \\
&& +\int \di Q \; (\slashed{p}_{3}+m)\Gamma_{2}(\slashed{p}_{2}+m)
 \left\{ \left[\Gamma_{1,\sigma}(p+q^{\prime}/2)i\partial_{q}^{\sigma}\right]
 \left[\e^{-iq^{0}(y_{\Sigma}^{0}-x^{0})}\delta^{3}({\bf q})\right]\right\} (\slashed{p}_{1}+m)
 \left[\e^{-iq^{\prime0}(y_{\Sigma}^{0}-x^{0})}\delta^{3}({\bf q}^{\prime})\right]\,,\\
&& W_{\text{quad,22}}(x,p) = \int \di Q \;(\slashed{p}_{3}+m)\Gamma_{2}(\slashed{p}_{2}+m)
\Gamma_{2}(\slashed{p}_{1}+m)\left[\e^{-iq^{\prime0}(y_{\Sigma}^{0}-x^{0})}\delta^{3}({\bf q}^{\prime})
\right]\left[\e^{-iq^{0}(y_{\Sigma}^{0}-x^{0})}\delta^{3}({\bf q})\right]\,.
\end{eqnarray*}
It is hard to obtain explicit formulae for the various $W_{\text{quad}}$, especially 
$W_{\text{quad,11}}$. Notwithstanding, calculating the axial-vector
component is relatively easy with the help of FeynCalc \cite{Shtabovenko:2023idz}.

We first focus on the first contribution to the $W_{\rm quad}$. Working out the trace of
Dirac matrices, we find:
\begin{eqnarray}\label{Wquad11}
\tr\left[\gamma^{\mu}\gamma^{5}W_{\text{quad,11}}(x,p)\right] & = & 
8p^{0}\epsilon^{\mu\nu\alpha\rho}p_{\nu}\int \di Q \;\left\{ \hat{t}_{\rho}(q_{\alpha}+
q_{\alpha}^{\prime})-\frac{1}{2p^{0}}\left[q_{\alpha}q_{\rho}^{\prime}+\hat{t}\cdot(q-q^{\prime})(q_{\alpha}+q_{\alpha}^{\prime})\hat{t}_{\rho}\right]\right\} \nonumber \\
 &  & \times\left[(p^{\nu^{\prime}}-q^{\nu^{\prime}}/2)\partial_{\lambda^{\prime}}
 \beta_{\nu^{\prime}}(x)-\partial_{\lambda^{\prime}}\zeta(x)\right]\left[(p^{\nu}+
 q^{\prime\nu}/2)\partial_{\lambda}\beta_{\nu}(x)-\partial_{\lambda}\zeta(x)\right]\nonumber \\
 &  & \times\left\{ \partial_{q}^{\lambda}\partial_{q^{\prime}}^{\lambda^{\prime}}
 \sin\left[(q^{0}+q^{\prime0})(y_{\Sigma}^{0}-x^{0})\right]\delta^{3}({\bf q}^{\prime})
 \delta^{3}({\bf q})\right\}\,,
\end{eqnarray}
taking advantage of the fact that \eqref{Wquad11} is a real function. Integrating by parts, 
we can rewrite it as:
\begin{equation}\label{eq:Tr-Wigner-c}
\tr\left[\gamma^{\mu}\gamma^{5}W_{\text{quad,11}}(x,p)\right]=\frac{4p^{0}}
{(2\pi)^{3}}n_{F}(p)[1-n_{F}(p)][1-2n_{F}(p)]\epsilon^{\mu\nu\alpha\rho}p_{\nu}(I_{\alpha\rho}^{(1)}+I_{\alpha\rho}^{(2)}+I_{\alpha\rho}^{(3)}+I_{\alpha\rho}^{(4)})\,,
\end{equation}
where:
\begin{eqnarray*}
I_{\alpha\rho}^{(1)} & \equiv & \int \di^{4}q \, \di^{4}q^{\prime}\; g_{1}(p;q,q^{\prime})g_{2}(p;q,q^{\prime})\left[\partial_{q}^{\lambda}\partial_{q^{\prime}}^{\lambda^{\prime}}
g_{3,\alpha\rho\lambda\lambda^{\prime}}(p;q,q^{\prime})\right]\,,\nonumber \\
I_{\alpha\rho}^{(2)} & \equiv & \int \di^{4}q \, \di^{4}q^{\prime} \; g_{1}(p;q,q^{\prime})\left[\partial_{q}^{\lambda}\partial_{q^{\prime}}^{\lambda^{\prime}}g_{2}(p;q,q^{\prime})\right]g_{3,\alpha\rho\lambda\lambda^{\prime}}(p;q,q^{\prime})\,,\nonumber \\
I_{\alpha\rho}^{(3)} & \equiv & \int \di^{4}q \,  \di^{4}q^{\prime} \; g_{1}(p;q,q^{\prime})
\left[\partial_{q}^{\lambda}g_{2}(p;q,q^{\prime})\right]
\left[\partial_{q^{\prime}}^{\lambda^{\prime}}g_{3,\alpha\rho\lambda\lambda^{\prime}}(p;q,q^{\prime})
\right]\,,\nonumber \\
I_{\alpha\rho}^{(4)} & \equiv & \int \di^{4}q \, \di^{4}q^{\prime} \; g_{1}(p;q,q^{\prime})
\left[\partial_{q^{\prime}}^{\lambda^{\prime}}g_{2}(p;q,q^{\prime})\right]\left[\partial_{q}^{\lambda}
g_{3,\alpha\rho\lambda\lambda^{\prime}} (p;q,q^{\prime})\right]\,,
\end{eqnarray*}
with auxiliary functions
\begin{eqnarray*}
g_{1}(p;q,q^{\prime}) & \equiv & \sin\left[(q^{0}+q^{\prime0})(y_{\Sigma}^{0}-x^{0})\right]
\delta^{3}({\bf q}^{\prime})\delta^{3}({\bf q})\text{sgn}(p_{1}^{0})\text{sgn}(p_{2}^{0})
\text{sgn}(p_{3}^{0})\,,\nonumber \\
g_{2}(p;q,q^{\prime}) & \equiv & \delta(p_{1}^{2}-m^{2})\delta(p_{2}^{2}-m^{2})\delta(p_{3}^{2}-m^{2})\,,\nonumber \\
g_{3,\alpha\rho\lambda\lambda^{\prime}}(p;q,q^{\prime}) & \equiv & 
\left[1+\frac{1-4\e^{\beta\cdot p}+\e^{2\beta\cdot p}}{6\left(\e^{2\beta\cdot p}-1\right)}
\beta\cdot(q-q^{\prime})\right]\nonumber \\
 &  & \times\left\{ \hat{t}_{\rho}(q_{\alpha}+q_{\alpha}^{\prime})-\frac{1}{2p^{0}}
 \left[q_{\alpha}q_{\rho}^{\prime}+\hat{t}\cdot(q-q^{\prime})(q_{\alpha}+q_{\alpha}^{\prime})
 \hat{t}_{\rho}\right]\right\} \nonumber \\
 &  & \times\left[(p^{\tau^{\prime}}-q^{\tau^{\prime}}/2)\partial_{\lambda^{\prime}}\beta_{\tau^{\prime}}(x)-\partial_{\lambda^{\prime}}\zeta(x)\right]\left[(p^{\tau}+q^{\prime\tau}/2)
 \partial_{\lambda}\beta_{\tau}(x)-\partial_{\lambda}\zeta(x)\right]\,.
\end{eqnarray*}
We note that we have neglected momentum derivatives of the sign functions because they 
will give terms like $\delta(p_{1}^{0})$, a constraint that in general cannot be fulfilled. 

In the term $I_{\alpha\rho}^{(1)}$, the delta functions in $g_{1}$ and $g_{2}$ yield:
\begin{equation}\label{eq:delta-functions}
\delta(p_{1}^{2}-m^{2})\delta(p_{2}^{2}-m^{2})\delta(p_{3}^{2}-m^{2})\delta^{3}({\bf q})
\delta^{3}({\bf q}^{\prime})=\frac{\delta(p^{2}-m^{2})}{4(p^{0})^{2}}\delta^{4}(q)\delta^{4}(q^{\prime})\,,
\end{equation}
forcing $q^{\mu}$ and $q^{\prime\mu}$ to be vanishing vectors. Therefore $I_{\alpha\rho}^{(1)}$ 
vanishes because:
\begin{equation*}
I_{\alpha\rho}^{(1)}\propto\lim_{q^{\mu},q^{\prime\mu}\rightarrow0}
\sin\left[(q^{0}+q^{\prime0})(y_{\Sigma}^{0}-x^{0})\right]=0\,.
\end{equation*}
In the term $I_{\alpha\rho}^{(2)}$, the momentum derivatives act on $g_{2}$, which can be 
expressed as:
\begin{equation*}
g_{2}(p;q,q^{\prime})=\frac{1}{4}\delta\left(p^{2}-m^{2}+\frac{q^{2}+q^{\prime2}}{4}+
\frac{q\cdot q^{\prime}}{2}\right)\delta\left[q\cdot\left(p+\frac{q^{\prime}}
{2}\right)\right]\delta\left[q^{\prime}\cdot\left(p-\frac{q}{2}\right)\right]\,.
\end{equation*}
If both $\partial_{q}^{\lambda}$ and $\partial_{q^{\prime}}^{\lambda^{\prime}}$ act on 
$\delta\left(p^{2}-m^{2}+\frac{q^{2}+q^{\prime2}}{4}+\frac{q\cdot q^{\prime}}{2}\right)$,
the remaining two delta functions in $g_{2}$, with the delta functions in $g_{1}$, will 
lead to $\delta^{4}(q)\delta^{4}(q^{\prime})$ and thus $g_{1}=0$ because:
$$
\lim_{q^{\mu},q^{\prime\mu}\rightarrow0}\sin\left[(q^{0}+q^{\prime0})(y_{\Sigma}^{0}-x^{0})\right]=0.
$$
If $\partial_{q}^{\lambda}$ or $\partial_{q^{\prime}}^{\lambda^{\prime}}$(or both of them) 
acts on $\delta\left[q\cdot\left(p+\frac{q^{\prime}}{2}\right)\right]$ or 
$\delta\left[q^{\prime}\cdot\left(p-\frac{q}{2}\right)\right]$, one can convert it to a 
derivative with respect to the time component $q^{0}$ or $q^{\prime0}$ and then integrate
by parts. The only term that survives in our calculation reads:
\begin{eqnarray*}
I_{\alpha\rho}^{(2)} & = & \frac{1}{4}\int \di^{4}q \di^{4}q^{\prime} \; g_{1}(p;q,q^{\prime})
\left(p^{\lambda}+\frac{q^{\prime\lambda}}{2}\right)\left(p^{\lambda^{\prime}}-
\frac{q^{\lambda^{\prime}}}{2}\right)g_{3,\alpha\rho\lambda\lambda^{\prime}}(p;q,q^{\prime})
\nonumber \\
 &  & \times\delta\left(p^{2}-m^{2}+\frac{q^{2}+q^{\prime2}}{4}+
 \frac{q\cdot q^{\prime}}{2}\right)\delta^{\prime}\left[q\cdot\left(p+\frac{q^{\prime}}{2}\right)\right]
 \delta^{\prime}\left[q^{\prime}\cdot\left(p-\frac{q}{2}\right)\right]\,.
\end{eqnarray*}
Integrating again by parts, we find:
\begin{equation*}
I_{\alpha\rho}^{(2)}\propto\lim_{q,q^{\prime}\rightarrow0}
\frac{\partial}{\partial q^{0}}\frac{\partial}{\partial q^{\prime0}}
\sin\left[(q^{0}+q^{\prime0})(y_{\Sigma}^{0}-x^{0})\right]\hat{t}_{\rho}
(q_{\alpha}+q_{\alpha}^{\prime})\propto\hat{t}_{\alpha}\hat{t}_{\rho}\,,
\end{equation*}
which is symmetric with respect to its indices $\alpha,\rho$. 

The calculation of $I_{\alpha\rho}^{(3)}$ is similar to that of $I_{\alpha\rho}^{(2)}$, 
and the only term that survive reads:
\begin{eqnarray*}
I_{\alpha\rho}^{(3)} & = & \frac{1}{4}p^{\lambda}\int \di^{4}q \, \di^{4}q^{\prime} \; 
g_{1}(p;q,q^{\prime})\delta\left(p^{2}-m^{2}+\frac{q^{2}+q^{\prime2}}{4}+
\frac{q\cdot q^{\prime}}{2}\right)\nonumber \\
 &  & \times\delta^{\prime}\left[q\cdot\left(p+\frac{q^{\prime}}{2}\right)\right]
 \delta\left[q^{\prime}\cdot\left(p-\frac{q}{2}\right)\right]
 \left[\partial_{q^{\prime}}^{\lambda^{\prime}}
 g_{3,\alpha\rho\lambda\lambda^{\prime}}(p;q,q^{\prime})\right]\,.
\end{eqnarray*}
Then, $\delta^{\prime}\left[q\cdot\left(p+\frac{q^{\prime}}{2}\right)\right]$
gives rise to a derivative with respect to $q^{0}$, and after integrating by parts, it turns out:
\begin{eqnarray*}
I_{\alpha\rho}^{(3)} & = & -\frac{1}{4}p^{\lambda}\lim_{q^{\mu},q^{\prime\mu}\rightarrow0}
\frac{\partial}{\partial q^{0}}\sin\left[(q^{0}+q^{\prime0})(y_{\Sigma}^{0}-x^{0})\right]
\text{sgn}(p_{1}^{0})\text{sgn}(p_{2}^{0})\text{sgn}(p_{3}^{0})\nonumber \\
 &  & \times\frac{\delta\left[p^{2}-m^{2}+\frac{q^{2}+q^{\prime2}}{4}+
 \frac{q\cdot q^{\prime}}{2}\right]}{\left(p^{0}+\frac{q^{\prime0}}{2}\right)\left|p^{0}-
 \frac{q^{0}}{2}\right|\left|p^{0}+\frac{q^{\prime0}}{2}\right|}\left[\partial_{q^{\prime}}^{\lambda^{\prime}}
 g_{3,\alpha\rho\lambda\lambda^{\prime}}(p;q,q^{\prime})\right]\nonumber \\
 & = & -\frac{\delta(p^{2}-m^{2})\text{sgn}(p^{0})}{4(p^{0})^{3}}(y_{\Sigma}^{0}-x^{0})\hat{t}_{\rho}g_{\alpha}^{\lambda^{\prime}}\left[p^{\tau^{\prime}}
 \partial_{\lambda^{\prime}}\beta_{\tau^{\prime}}(x)-\partial_{\lambda^{\prime}}\zeta(x)\right]
 p^{\lambda}\left[p^{\tau}\partial_{\lambda}\beta_{\tau}(x)-\partial_{\lambda}\zeta(x)\right]\,.
\end{eqnarray*}
Finally, the term $I_{\alpha\rho}^{(4)}$ yields an expression which turns out to
be the same as that of $I_{\alpha\rho}^{(3)}$ because of the symmetry in the exchange 
of $q^{\mu}$ with $q^{\prime\mu}$. Substituting $I_{\alpha\rho}^{(i)}$ with $i=1,2,3,4$ 
back into eq. (\ref{eq:Tr-Wigner-c}), we obtain:
\begin{eqnarray*}
\tr\left[\gamma^{\mu}\gamma^{5}W_{\text{quad,11}}(x,p)\right] & = & 
-\frac{2\delta(p^{2}-m^{2})\text{sgn}(p^{0})}{(2\pi)^{3}(p^{0})^{2}}n_{F}(p)
[1-n_{F}(p)][1-2n_{F}(p)](y_{\Sigma}^{0}-x^{0})\nonumber \\
 &  & \times\epsilon^{\mu\nu\alpha\rho}p_{\nu}\hat{t}_{\rho}g_{\alpha}^{\lambda}
 \left[p^{\tau}\partial_{\lambda}\beta_{\tau}(x)-\partial_{\lambda}\zeta(x)\right]
 p^{\lambda^{\prime}}\left[p^{\tau^{\prime}}
 \partial_{\lambda^{\prime}}\beta_{\tau^{\prime}}(x)-\partial_{\lambda^{\prime}}
 \zeta(x)\right]\,.
\end{eqnarray*}

Now we can proceed to the calculation of $W_{\text{quad},12}$. First, we rewrite
the delta functions to the following form:
\begin{eqnarray*}
\delta(p_{1}^{2}-m^{2})\delta(p_{2}^{2}-m^{2})\delta(p_{3}^{2}-m^{2})\delta^{3}({\bf q}) 
& = & \frac{1}{4\left|p^{0}+\frac{q^{\prime0}}{2}\right|}\delta\left[p^{2}+
\frac{(q^{\prime})^{2}}{4}-m^{2}\right]\delta(p\cdot q^{\prime})\delta^{(4)}(q)\,,\nonumber \\
\delta(p_{1}^{2}-m^{2})\delta(p_{2}^{2}-m^{2})\delta(p_{3}^{2}-m^{2})
\delta^{3}({\bf q}^{\prime}) & = & \frac{1}{4\left|p^{0}-\frac{q^{0}}{2}\right|}
\delta\left[p^{2}+\frac{q^{2}}{4}-m^{2}\right]\delta(p\cdot q)\delta^{(4)}(q^{\prime})\,.
\end{eqnarray*}
Plugging them into $W_{\text{quad},12}$ and integrating
out $q^{\mu}$ or $q^{\prime\mu}$:
\begin{eqnarray*}
W_{\text{quad,12}}(x,p) & = & \frac{1}{32(2\pi)^{3}}\int \di^{4}q
\delta\left[p^{2}+\frac{q^{2}}{4}-m^{2}\right]\delta(p\cdot q)n_{F}(p)
[1-n_{F}(p)][1-2n_{F}(p)]\nonumber\\
&&\times\text{sgn}\left(p^{0}-\frac{q^{0}}{2}\right)\text{sgn}\left(p^{0}+
\frac{q^{0}}{2}\right)[p^{\nu}\partial_{\lambda}\beta_{\nu}(x)-
\partial_{\lambda}\zeta(x)]\left[\gamma\cdot(p-q/2)+m\right]\nonumber \\
 &  & \times\left\{\frac{1}{p^{0}+\frac{q^{0}}{2}}\left[1-\frac{1-4\e^{\beta\cdot p}
 +\e^{2\beta\cdot p}}{6\left(\e^{2\beta\cdot p}-1\right)}\beta\cdot q\right](\hat{t}\cdot\gamma)\left[\gamma\cdot(p+q/2)+m\right]\gamma_{\sigma}\gamma^{5}\right.\nonumber \\
 &  & \left.+\frac{1}{p^{0}-\frac{q^{0}}{2}}\left[1+\frac{1-4\e^{\beta\cdot p}
 +\e^{2\beta\cdot p}}{6\left(\e^{2\beta\cdot p}-1\right)}\beta\cdot q\right]
 \gamma_{\sigma}\gamma^{5}\left[\gamma\cdot(p-q/2)+m\right](\hat{t}\cdot\gamma)\right\}\nonumber \\
 &  & \times\left[\gamma\cdot(p+q/2)+m\right]\hat{t}_{\tau}\epsilon^{\tau\nu\alpha\sigma}
 \Omega_{\nu\alpha}(x)\left\{ i\partial_{q}^{\lambda}\left[\e^{-iq^{0}(y_{\Sigma}^{0}-x^{0})}
 \delta^{3}({\bf q})\right]\right\}\,.
\end{eqnarray*}
Integrating by parts, we can rewrite it as follows:
\begin{eqnarray*}
W_{\text{quad,12}}(x,p) & = & \frac{\delta(p^{2}-m^{2})}{32(2\pi)^{3}|p^{0}|}
n_{F}(p)[1-n_{F}(p)][1-2n_{F}(p)][p^{\nu}\partial_{\lambda}\beta_{\nu}(x)-
\partial_{\lambda}\zeta(x)]\hat{t}_{\tau}\epsilon^{\tau\nu\alpha\sigma}\Omega_{\nu\alpha}(x)\nonumber \\
 &  & \times\lim_{q\rightarrow0}D_{q}^{\lambda}\left[\gamma\cdot(p-q/2)+m\right]
 \left\{\frac{1}{p^{0}+\frac{q^{0}}{2}}\left[1-\frac{1-4\e^{\beta\cdot p}+
 \e^{2\beta\cdot p}}{6\left(\e^{2\beta\cdot p}-1\right)}\beta\cdot q\right]\hat{t}\cdot\gamma)\left[\gamma\cdot(p+q/2)+m\right]\gamma_{\sigma}\gamma^{5}\right.\nonumber \\
 &  & \left.+\frac{1}{p^{0}-\frac{q^{0}}{2}}\left[1+\frac{1-4\e^{\beta\cdot p}+
 \e^{2\beta\cdot p}}{6\left(\e^{2\beta\cdot p}-1\right)}\beta\cdot q\right]
 \gamma_{\sigma}\gamma^{5}\left[\gamma\cdot(p-q/2)+m\right](\hat{t}\cdot\gamma)\right\}
 \left[\gamma\cdot(p+q/2)+m\right]\,,
\end{eqnarray*}
where the operator $D_{q}^{\lambda}$ is given in eq. (\ref{operator-D}). The axial-vector
component is given by:
\begin{eqnarray*}
\tr\left[\gamma^{\mu}\gamma^{5}W_{\text{quad,12}}(x,p)\right] & = & 
-\frac{\delta(p^{2}-m^{2})\text{sgn}(p^{0})}{(2\pi)^{3}(p^{0})^{2}}n_{F}(p)
[1-n_{F}(p)][1-2n_{F}(p)](y_{\Sigma}^{0}-x^{0})\nonumber \\
 &  & \times p^{\lambda}[p^{\nu}\partial_{\lambda}\beta_{\nu}(x)-\partial_{\lambda}
 \zeta(x)](g_{\sigma}^{\mu}m^{2}-p^{\mu}p_{\sigma})\hat{t}_{\tau}
 \epsilon^{\tau\nu\alpha\sigma}\Omega_{\nu\alpha}(x)\,.
\end{eqnarray*}
The calculation of $W_{\text{quad,22}}$ is quite straightforward by using the relation 
(\ref{eq:delta-functions}). Integrating out the delta functions for $q^{\mu}$ and $q^{\prime\mu}$, 
we arrive at:
\begin{eqnarray*}
W_{\text{quad,22}}(x,p) & = & \frac{\delta(p^{2}-m^{2})
\text{sgn}(p^{0})}{8(2\pi)^{3}(p^{0})^{2}}n_{F}(p)[1-n_{F}(p)][1-2n_{F}(p)]\nonumber \\
 &  & \times(\slashed{p}+m)\Gamma_{2}(\slashed{p}+m)\Gamma_{2}(\slashed{p}+m)\,,
\end{eqnarray*}
which has a vanishing axial-vector component:
\begin{equation*}
\tr\left[\gamma^{\mu}\gamma^{5}W_{\text{quad,22}}(x,p)\right]=0\,.
\end{equation*}
Hence, the whole quadratic contribution to axial-vector component of the Wigner function 
at the second order reads:
\begin{eqnarray}\label{wquadlast}
\tr\left[\gamma^{\mu}\gamma^{5}W_{\text{quad}}(x,p)\right] & = & \frac{\delta(p^{2}-m^{2})\text{sgn}(p^{0})}{(2\pi)^{3}(p^{0})^{2}}n_{F}(p)[1-n_{F}(p)][1-2n_{F}(p)](y_{\Sigma}^{0}-x^{0})\nonumber \\
 &  & \times p^{\lambda^{\prime}}\partial_{\lambda^{\prime}}\left[p^{\tau^{\prime}}\beta_{\tau^{\prime}}(x)-\zeta(x)\right]\nonumber \\
 &  & \times\left\{ 2\epsilon^{\mu\nu\rho\lambda}p_{\nu}\hat{t}_{\rho}\partial_{\lambda}\left[p^{\tau}\beta_{\tau}(x)-\zeta(x)\right]+(p^{\mu}p_{\tau}-g_{\tau}^{\mu}m^{2})\hat{t}_{\rho}\epsilon^{\rho\nu\lambda\tau}\Omega_{\nu\lambda}(x)\right\}\,.
\end{eqnarray}
Comparing eq. \eqref{wquadlast} with eqs. (\ref{W-indep-final}) and (\ref{Wigner-components-1}) 
one can finally prove eq. (\ref{Wigner-quad-2}).

\end{document}